\pdfoutput=1

\documentclass[11pt,twoside,a4paper,cmspaper,final,collab]{cms-tdr}
\def\svnVersion{490875P}\def\cmsCernNoTag{CERN-EP-2018-141}\def\cmsCernDate{\today}\def\cmsMessage{Published in Physics Letters B as \href{http://dx.doi.org/10.1016/j.physletb.2018.12.073}{\doi{10.1016/j.physletb.2018.12.073.}}}

\begin{document}\cmsNoteHeader{HIG-16-042}

\hyphenation{had-ron-i-za-tion}
\hyphenation{cal-or-i-me-ter}
\hyphenation{de-vices}
\RCS$HeadURL: svn+ssh://svn.cern.ch/reps/tdr2/papers/HIG-16-042/trunk/HIG-16-042.tex $
\RCS$Id: HIG-16-042.tex 483066 2018-12-01 14:13:23Z alverson $

\newlength\cmsFigWidth
\ifthenelse{\boolean{cms@external}}{\setlength\cmsFigWidth{0.85\columnwidth}}{\setlength\cmsFigWidth{0.4\textwidth}}

\newlength\cmsFigWidthAlt
\ifthenelse{\boolean{cms@external}}{\setlength\cmsFigWidthAlt{0.98\columnwidth}}{\setlength\cmsFigWidthAlt{0.48\textwidth}}

\newlength\cmsFigWidthSingle
\ifthenelse{\boolean{cms@external}}{\setlength\cmsFigWidthSingle{0.98\columnwidth}}{\setlength\cmsFigWidthSingle{0.7\textwidth}}

\ifthenelse{\boolean{cms@external}}{\providecommand{\cmsLeft}{top\xspace}}{\providecommand{\cmsLeft}{left\xspace}}
\ifthenelse{\boolean{cms@external}}{\providecommand{\cmsRight}{bottom\xspace}}{\providecommand{\cmsRight}{right\xspace}}

\ifthenelse{\boolean{cms@external}}{\providecommand{\cmsTable}[1]{#1}}{\providecommand{\cmsTable}[1]{\resizebox{\textwidth}{!}{#1}}}

\newlength\cmsTabSkip\setlength{\cmsTabSkip}{1ex}
\newlength\extraTabSkip\setlength{\extraTabSkip}{2ex}

\providecommand{\mt}{\ensuremath{m_\mathrm{T}}\xspace}
\newcommand{\ptll}{\ensuremath{\pt^{\ell\ell}}\xspace}
\newcommand{\mll}{\ensuremath{m_{\ell\ell}}\xspace}
\newcommand{\delphillmet}{\ensuremath{\Delta\phi(\ell\ell,\ptvecmiss)}}
\newcommand{\delphiltwomet}{\ensuremath{\Delta\phi(\ell 2,\ptvecmiss)}}
\newcommand{\mllminlll}{\ensuremath{\text{min--}m_{\ell^+\ell^-}}\xspace}
\newcommand{\mjj}{\ensuremath{m_{jj}}\xspace}
\newcommand{\detajj}{\ensuremath{\Delta\eta_{jj}}}
\newcommand{\mtwTwo}{\ensuremath{\mt^{\ell 2,\ptmiss}}\xspace}
\newcommand{\wjets}{\ensuremath{\PW{+}\text{jets}}\xspace}

\hyphenation{ATLAS}

\cmsNoteHeader{HIG-16-042}
\title{Measurements of properties of the Higgs boson decaying to a $\PW$ boson pair in $\Pp\Pp$ collisions at \texorpdfstring{$\sqrt{s}=13\TeV$}}

\date{\today}

\abstract{
Measurements of the production of the standard model Higgs boson decaying to a $\PW$ boson pair are reported.
The $\PWp\PWm$ candidates are selected in events with an oppositely charged lepton pair, large missing transverse momentum, and various numbers of jets. To select Higgs bosons
produced via vector boson fusion and associated production with a $\PW$ or $\PZ$ boson, events with two jets or three or four leptons are also selected. The event sample corresponds to an integrated luminosity of $35.9\fbinv$, collected in $\Pp\Pp$ collisions at $\sqrt{s} = 13\TeV$ by the CMS detector at the LHC during 2016. Combining all channels, the observed cross section times branching fraction is $1.28 ^{+0.18}_{-0.17}$ times the standard model prediction for the Higgs boson with a mass of $125.09\GeV$.
This is the first observation of the Higgs boson decay to $\PW$ boson pairs by the CMS experiment.
}

\hypersetup{
pdfauthor={CMS Collaboration},%
pdftitle={Measurements of properties of the Higgs boson decaying to a W boson pair in pp collisions at sqrt(s)=13 TeV},%
pdfsubject={CMS},%
pdfkeywords={Higgs, WW}}

\maketitle

\section{Introduction \label{section:intro} }

In the standard model (SM) of particle physics, the origin of the masses of the $\PW$ and $\PZ$ bosons is based on the spontaneous breaking of the electroweak symmetry.
This symmetry breaking is achieved through the introduction of a complex doublet scalar
field~\cite{Higgs1,Higgs2,Higgs3,Higgs4,Higgs5,Higgs6}, leading to the prediction of the existence of one physical neutral scalar particle,
commonly known as the Higgs boson ($\PH$). The observation of a new particle at a mass of approximately 125\GeV with Higgs boson-like
properties was reported by the ATLAS~\cite{AtlasPaperCombination}
and CMS~\cite{CMSPaperCombination,Chatrchyan:2013lba} Collaborations during the first running period of the CERN LHC in proton-proton ($\Pp\Pp$) collisions at center-of-mass energies of 7 and 8\TeV.
Subsequent publications from both collaborations, based on the 7 and 8\TeV data sets~\cite{Aad:2015gba,Aad:2013xqa,CMS_combination,Khachatryan:2014kca},
established that all measured properties of the new particle, including its spin, parity,
and coupling strengths to SM particles, are consistent within the uncertainties with those expected for the SM Higgs boson.
A combination of the ATLAS and CMS results~\cite{Aad:2015zhl,Khachatryan:2016vau}
further confirmed these observations and resulted in determining the boson mass
to be $m_\PH = 125.09 \pm 0.21\stat \pm 0.11\syst\GeV$.

The Higgs boson decay to a pair of $\PW$ bosons was studied by the ATLAS and CMS Collaborations using the 7 and 8\TeV data sets
in leptonic final states, exploring several production mechanisms~\cite{ATLAS:2014aga,Aad:2015ona,Chatrchyan:2013iaa}.
The probability of observing a signal at least as large as the one seen, under
the background-only hypothesis, corresponded to a significance of 6.5 and 4.3
standard deviations (s.d.) for ATLAS and CMS respectively, while the expected
significance for a SM Higgs boson was 5.8 (5.9) s.d. for the CMS (ATLAS) collaboration.
A later CMS combination~\cite{CMS_combination}, that includes Higgs boson
production in association with a top quark pair, reported an observed significance of 4.7 s.d. for this decay. The same decay channel was used by the ATLAS and CMS Collaborations to search for the Higgs boson off-shell production~\cite{Aad:2015xua,Khachatryan:2016ctc} and to perform fiducial and differential cross section measurements~\cite{Aad:2016lvc,Khachatryan:2016vnn}.

In 2015, the LHC restarted at $\sqrt{s}=13\TeV$, delivering high luminosity $\Pp\Pp$ collisions.
The new data are used to further constrain the properties of the Higgs boson:
any significant deviation from the SM predictions would be a clear sign of new physics.
This paper presents the analysis of the $\PH\to\PW\PW$ decay at 13\TeV,
using a data sample corresponding to a total integrated luminosity of $35.9\fbinv$, collected during 2016.
The same final state was recently studied by ATLAS~\cite{Aaboud:2018jqu} using
2015 and 2016 data.

Gluon fusion ($\Pg\Pg\PH$) is the dominant production mode for a Higgs boson with a mass of 125\GeV in $\Pp\Pp$ collisions at $\sqrt{s}=13\TeV$.
The large Higgs boson branching fraction to a $\PW$ boson pair makes this channel
suitable for a precision measurement of the Higgs boson production cross section, and also allows studies of subleading production channels, such as Higgs boson production via vector boson fusion (VBF) and associated
production with a vector boson (V$\PH$). These channels are also studied in this paper, contributing to the precision in the measurement of the Higgs boson couplings.

The leptonic decays of the two $\PW$ bosons provide the cleanest decay channel, despite the presence of neutrinos in the final state that prevents the full reconstruction of the Higgs boson mass. The different-flavor (DF) leptonic decay mode $\Pe\PGm$ has the largest branching fraction, is the least affected
by background processes, and therefore is the most sensitive channel of the analysis.
The same-flavor (SF) $\EE$ and
$\MM$ final states are also considered, although their sensitivity is
limited by the contamination from the Drell--Yan (DY) background with missing transverse momentum due to instrumental effects.

Events with a pair of oppositely charged leptons (electrons and/or muons)
and missing transverse momentum, due to the presence of neutrinos in the final state, are selected.
This signature is common to other SM processes that contribute to the background in this analysis. The main contribution comes from nonresonant production
of $\PW$ boson pairs ($\PW\PW$), an irreducible
background that shares the same final state and can only be separated from the signal using kinematic distributions. Backgrounds coming from top quark events (\ttbar and t$\PW$) are also important,
followed by other processes, such as \wjets and other diboson and triboson production processes.
The DY process is the dominant source of background in the dielectron and dimuon
final states, while it is subdominant in the electron-muon final state, since its
contribution arises from the leptonic decays of the $\tau$ leptons emerging from
$\PZ/\Pgg^{*}\to\TT$.

The events are categorized by jet multiplicity to better handle the \ttbar
background. In addition, dedicated categories are designed to enhance the sensitivity to the
VBF and V$\PH$ production mechanisms.

\section{The CMS detector \label{section:cms}}

The CMS detector is a multipurpose apparatus designed to study high transverse
momentum ($\pt$) physics processes in proton-proton and heavy ion collisions,
and is described in detail in Ref.~\cite{CMSdetector} together with a definition
of the coordinate system used. A superconducting solenoid occupies its central
region, providing a magnetic field of 3.8\unit{T} parallel to the beam direction.
Charged particle trajectories are measured by the silicon pixel and strip trackers,
which cover a pseudorapidity region of $\abs{\eta} < 2.5$. A lead tungstate crystal electromagnetic
calorimeter (ECAL), and a brass and scintillator hadron calorimeter surround the tracking
volume and cover $\abs{\eta} < 3$. The steel and quartz fiber Cherenkov hadron forward
calorimeter extends the coverage to $\abs{\eta} < 5$. The muon system consists
of gas-ionization detectors embedded in the steel flux-return yoke outside the
solenoid, and covers $\abs{\eta} < 2.4$. The first level of the CMS trigger system,
composed of custom hardware processors, is designed to select the most interesting
events in less than 4\mus, using information from the calorimeters and muon detectors.
The high-level trigger processor farm further reduces the event rate to about 1000\unit{Hz}, before data storage.

\section{Data and simulated samples \label{section:mcdata}}

The events used in this analysis are selected by high-level trigger algorithms that require the presence of one or two
high-\pt electrons or muons passing loose identification and isolation requirements. In single-lepton triggers, relatively tight lepton
identification criteria are applied.
The \pt threshold is $25\GeV$ in the central region
($\abs{\eta} < 2.1$) and $27\GeV$ for $2.1 < \abs{\eta} < 2.5$ for electrons,
while it is $24\GeV$ for muons ($\abs{\eta} < 2.4$).
In the dielectron trigger, the minimum required \pt is $23\GeV$
for the leading and $12\GeV$ for the subleading electron.
In the dimuon trigger, the minimum \pt is $17\GeV$ for the leading
and $8\GeV$ for the subleading muon.
In the two dilepton e$\mu$ triggers used in the analysis, the minimum
\pt requirements are either $8\GeV$ for the muon and $23\GeV$ for the electron,
or $23\GeV$ for the muon and $12\GeV$ for the electron.
The combination of single-lepton and dilepton triggers provides an overall
trigger efficiency in excess of 98\% for selected signal events.

Several event generators are used to optimize the analysis and estimate
the expected yields of signal and backgrounds, as well as their associated systematic uncertainties.
Different Higgs boson production mechanisms are simulated. Both $\Pg\Pg\PH$ and VBF
are generated with \POWHEG{} v2~\cite{Nason:2004rx,Frixione:2007vw,Alioli:2010xd,Bagnaschi:2011tu},
which describes the full next-to-leading order (NLO) perturbative quantum chromodynamics (QCD) properties of these
processes. In addition, the $\Pg\Pg\PH$ process is reweighted to match the
Higgs boson \pt and the number of associated jets to the prediction of \POWHEG{} \textsc{nnlops}~\cite{Hamilton:2013fea},
which provides a next-to-next-to-leading order (NNLO) description for the inclusive Higgs boson production, NLO for the exclusive $\PH+1$ jet production, and leading order (LO) for the exclusive $\PH+2$ jets production. The reweighting is performed by computing the ratio of the Higgs boson \pt distribution from the \textsc{nnlops} generator to that from the \POWHEG{} generator in each jet multiplicity bin, and applying this ratio to the $\Pg\Pg\PH$ \POWHEG{} simulation.
The \textsc{minlo hvj}~\cite{Luisoni:2013kna} extension of \POWHEG{} is used to simulate the associated production of the Higgs boson with vector bosons ($\PWp\PH$, $\PWm\PH$, $\PZ\PH$), which simulates the V$\PH+0$ and 1 jet processes with NLO accuracy.
Higgs boson production in association with top or bottom quarks, such as $\ttbar\PH$ and $\bbbar\PH$ production mechanisms, are considered as well, although they only contribute
to a minor extent in the phase space selected by this analysis. For the
simulation of $\ttbar\PH$ production the \POWHEG{} generator is used, while
the \MGvATNLO~v2.2.2 generator~\cite{Alwall:2014hca} is used to simulate the
$\bbbar\PH$ production. The Higgs boson is generated with a mass of
$125.09\GeV$ and is made to decay into a pair of $\PW$ bosons, considering only leptonic $\PW$ boson decays ($\Pe$, $\Pgm$, or $\PGt$).
For Higgs bosons produced via $\Pg\Pg\PH$~\cite{Alioli:2008tz} and VBF~\cite{Nason:2009ai} processes, their decay into two $\PW$ bosons and subsequently into leptons is simulated using
\textsc{jhugen}~v5.2.5~\cite{Gao:2010qx,Bolognesi:2012mm}.
For the associated production mechanisms, including gluon fusion
produced $\PZ\PH$, the Higgs boson decay and the associated vector boson
inclusive decays are simulated by \PYTHIA~8.212~\cite{Sjostrand:2007gs}.
The simulated signal samples are normalized using cross sections~\cite{deFlorian:2016spz} and decay rates~\cite{Heinemeyer:2013tqa}
computed by the LHC Higgs Cross Section Working Group. In particular the most recent next-to-next-to-next-to-leading order calculations for the
inclusive gluon fusion production are used~\cite{deFlorian:2016spz}.
Additional simulated samples, where the Higgs boson decays into a pair of
$\PGt$ leptons, are also produced for each of the aforementioned production
mechanisms. Unless stated otherwise, the $\PH\to\PGt\PGt$ events passing the selection are considered
signal events in the signal yield determination. However, their expected contribution
in the signal phase space is small compared to $\PH\to\PWp\PWm$.

The various background processes in this study are simulated as follows: \POWHEG v2~\cite{Melia:2011tj}
is used for $\qqbar\to\PW\PW$ production,
whereas $\cPg\cPg\to\PW\PW$ production is generated using \MCFM v7.0~\cite{Campbell:2013wga}. A $\PW\PW$ simulation with two additional jets is generated with \MGvATNLO at LO accuracy via diagrams with six electroweak (EW) vertices, referred to as $\PW\PW$ EW production.
In order to suppress the top quark background processes, the analysis is performed defining event categories with different number of high-\pt jets ($\pt > 30\GeV$). The classification of the events in bins of jet multiplicity spoils the convergence of
fixed-order calculations of the $\qqbar \to \PW\PW$ process and requires the use of dedicated resummation techniques for an
accurate prediction of the differential distributions~\cite{Meade:2014fca,Jaiswal:2014yba}.
The simulated $\qqbar \to \PW\PW$ events are therefore reweighted
to reproduce the $\pt^{\PW\PW}$ distribution from the \pt-resummed calculation.

The LO cross section for the $\cPg\cPg\to\PW\PW$ process is obtained directly from \MCFM. For this process,
the difference between LO and NLO cross sections is significant; a $K$ factor of 1.4 is calculated~\cite{Caola:2015rqy}
and applied to the $\cPg\cPg\to\PW\PW$ simulation. Given the theoretical uncertainties in the $K$ factor, and that it is mildly sensitive to the invariant mass of the $\PW\PW$ system ($m_{\PW\PW}$) in the phase space of interest, an $m_{\PW\PW}$-independent calculation is used.

Single top quark and \ttbar processes are generated using \POWHEG v2.
The cross sections of the different single top quark processes are estimated at NLO accuracy~\cite{Kant:2014oha}, while the \ttbar cross section is computed at NNLO accuracy, with next-to-next-to-leading-logarithmic soft-gluon resummation~\cite{Czakon:2011xx}.

The DY production of $\PZ/\Pgg^{*}$ is generated using \MGvATNLO at NLO accuracy using the FxFx jet matching and merging scheme with a merging scale $\mu{_\mathrm{Q}} = 30\GeV$~\cite{Frederix:2012ps},
and the $\PZ/\Pgg^{*}$ \pt distribution reweighted to match the distribution observed in data in dimuon events.

The $\PW\Pgg^{*}$ background was simulated with \POWHEG at NLO accuracy, down
to a minimum invariant mass of the virtual photon of 100~\MeV.
The effect of the $\Pgg^{*}$ mass cutoff was
estimated with a \MGvATNLO $\PW\Pgg$ LO sample, in which the photon pair
production was simulated by \PYTHIA in the parton shower approximation. The impact from
events in which the $\Pgg^{*}$ mass is below  100~\MeV was found to be one
order of magnitude smaller than the uncertainties quoted in this analysis,
thus their contribution was neglected.

Other multiboson processes, such as $\PW\PZ$, $\PZ\PZ$, and VVV ($\mathrm{V=\PW,\PZ}$), are also simulated with \MGvATNLO at NLO accuracy.

All processes are generated using the NNPDF~3.0~\cite{Ball:2013hta,Ball:2011uy} parton distribution functions (PDFs), with the accuracy matching that of the matrix element calculations. All the event generators are interfaced
to \PYTHIA for the showering of partons and hadronization, as well as the simulation of the underlying event (UE)
and multiple-parton interactions based on the CUET8PM1 tune~\cite{Khachatryan:2015pea}.

To estimate the systematic uncertainties related to the choice of the UE and multiple-parton interactions tune, the signal processes and the $\PW\PW$ background are also generated with alternative tunes, which are representative of the uncertainties in the CUET8PM1 tuning parameters.
The systematic uncertainty associated with showering and hadronization is estimated by interfacing the same samples with the
\HERWIG{}++ 2.7 generator~\cite{Richardson:2013nfo,Bellm:2013hwb}, using the UE-EE-5C tune for the simulation of UE and multiple-parton interactions~\cite{Khachatryan:2015pea}.

For all processes, the detector response is simulated using a detailed
description of the CMS detector, based on the \GEANTfour package~\cite{Agostinelli:2002hh}.
Additional simulated minimum bias $\Pp\Pp$ interactions
from \PYTHIA are overlapped with the event of interest in each collision to reproduce the number
of interactions per bunch crossing (pileup) measured in data.
The average number of pileup interactions is about 27 per event for the 2016 data set used in this analysis.

\section{Analysis strategy\label{eventsel}}

A particle-flow (PF) algorithm~\cite{Sirunyan:2017ulk} is used to reconstruct
the observable particles in the event. Energy deposits (clusters) measured by
the calorimeters and charged particle tracks identified in the central tracking
system and the muon detectors are combined to reconstruct individual particles.

Among the vertices reconstructed in the event, the one with the largest value
of summed physics-object $\pt^2$ is taken to be the primary $\Pp\Pp$
interaction vertex. The physics objects include those returned by a jet-finding algorithm~\cite{Cacciari:2008gp,Cacciari:2011ma} applied to all
charged tracks assigned to the vertex, and the associated missing transverse
momentum, defined as the negative vector sum of the \pt of those objects.

Electrons are reconstructed by matching clusters in the ECAL to tracks in the silicon tracker~\cite{Khachatryan:2015hwa}. In this analysis, electron candidates are required to have
$\abs{\eta}<2.5$. Additional requirements are applied to reject electrons originating from photon
conversions in the tracker material or jets misreconstructed as electrons.
Electron identification criteria rely on observables sensitive to the bremsstrahlung along the
electron trajectory, the geometrical and momentum-energy matching between the electron track
and the associated energy cluster in the ECAL, as well as ECAL shower shape observables and association with the primary vertex.

Muon candidates are reconstructed in the geometrical acceptance $\abs{\eta}<2.4$ by combining information from the silicon tracker and the muon system. Identification criteria based on the number of measurements in the
tracker and in the muon system, the fit quality of the muon track, and its consistency with its
origin from the primary vertex are imposed on the muon candidates to reduce the misidentification
rate.

Prompt leptons coming from EW interactions are usually isolated, whereas misidentified leptons and leptons
coming from jets are often accompanied by charged or neutral particles, and can arise from a secondary vertex.
Hence charged leptons are required to satisfy the isolation criterion that the \pt sum over charged PF candidates associated with the primary vertex, exclusive of the lepton itself, and neutral PF particles in a cone of a radius $\Delta R = \sqrt{\smash[b]{(\Delta\phi)^2+(\Delta\eta)^2}} = 0.4$ (0.3), where $\phi$ is the azimuthal angle in radians, centered on the muon (electron) direction is below a threshold of 15 (6)\% relative to the muon (electron) \pt. To mitigate the effect of the pileup on this isolation variable,
a correction based on the average energy density in the event~\cite{Cacciari:2007fd} is applied.
Additional requirements on the transverse ($\abs{d_{xy}}$) and longitudinal ($\abs{d_z}$) impact parameters with respect to the primary vertex are included.
Electrons detected by the ECAL barrel are required to have $\abs{d_z} < 0.10\cm$ and $\abs{d_{xy}} < 0.05\cm$, while electrons in the ECAL endcap must satisfy $\abs{d_z} < 0.20\cm$ and $\abs{d_{xy}} < 0.10\cm$.
For muons, the $\abs{d_z}$ parameter is required to be less than $0.10\cm$, while $\abs{d_{xy}}$ is required to be less than $0.01\cm$ for muons with $\pt<20\GeV$ and less than $0.02\cm$ for $\pt>20\GeV$.

The jet reconstruction starts with all PF candidates, and removes the charged ones that are not
associated with the primary vertex to mitigate the pileup impact.
The remaining charged PF candidates and all neutral candidates are clustered by the
anti-\kt algorithm~\cite{Cacciari:2008gp} with a distance parameter of 0.4.
To reduce further the residual pileup contamination from neutral PF candidates,
a correction based on the jet area~\cite{Cacciari:2007fd} is applied.
The jet energy is calibrated using both simulation and data following the technique described in Ref.~\cite{Chatrchyan:2011ds}.
To identify jets coming from $\PQb$ quarks ($\PQb$ jets), a multivariate (MVA) $\PQb$ tagging algorithm is used~\cite{Sirunyan:2017ezt}.
In this analysis, the chosen working point corresponds to about 80\% efficiency for genuine $\PQb$ jets, and to a mistagging rate of about 10\% for light-quark or gluon jets and of 35 to 50\% for $\PQc$ jets. A per-jet scale factor is computed and applied to account for $\PQb$ tagging
efficiency and mistagging rate differences between data and
simulation.

The missing transverse momentum vector (\ptvecmiss), whose magnitude is denoted as \ptmiss, is reconstructed as the negative
vectorial sum in the transverse plane of all PF particle candidate momenta.
Since the presence of pileup induces a degradation of the \ptmiss measurement, affecting mostly backgrounds with no genuine \ptmiss, such as DY production,
another \ptmiss that is constructed from only the charged particles (track \ptmiss) is used in events with an SF lepton pair ($\Pe\Pe$ or $\Pgm\Pgm$).
To suppress the remaining off-peak DY contribution in categories containing events with an SF lepton pair,
a dedicated MVA selection based on a boosted decision tree algorithm (BDT) is used, combining variables related to lepton kinematics and
\ptvecmiss.
The BDT is trained on simulated samples separately for different
jet multiplicity categories, and the output discriminator is used to define a phase space enriched in signal events and reduced DY background contamination.

Events are required to pass the single-lepton or dilepton triggers. For each event, this analysis requires at least two high-\pt lepton candidates with opposite sign, originating from the primary vertex, categorized as dielectron, dimuon, or $\Pe\Pgm$ pairs. Only jets with $\pt > 30$\GeV (20\GeV for $\PQb$ jets) and $\abs{\eta}<4.7$ ($\abs{\eta}<2.4$ for $\PQb$ jets) are considered in the analysis.
Jets are ignored if they overlap with an isolated lepton within a distance of $\Delta R = 0.3$. In addition,
the following kinematic selection is applied in the $\Pe\Pgm$ final state: one
electron and one muon are required to be reconstructed in the event with a
minimum \pt of 13\GeV for the electron and 10\GeV for the muon,
the higher \pt threshold for the electron resulting from the trigger
definition. One of the two leptons should also have a \pt greater than 25\GeV.
In the case of SF $\EE$ and $\MM$ final states, the leading lepton is required to have \pt greater than 25\GeV when it is an electron,
or 20\GeV when it is a muon. The subleading electron is required to have \pt greater
than 13\GeV, while for the muon a minimum \pt of 10\GeV is required.
Both leptons are required to be well identified, isolated, and prompt.

Given the large background contribution from \ttbar production in both DF and SF final states,
events are further categorized based on the number of jets in the event,
with the 0-jet category driving the sensitivity of the analysis.
A categorization of the selected events is performed, targeting different production mechanisms and different flavor compositions of the $\PW\PW$ decay products.

\section{Analysis categories}

\subsection{Different-flavor \texorpdfstring{$\Pg\Pg\PH$}{ggH} categories}\label{sec:AnalysisStrategyGGHOF}

The categories described in this section target the $\Pg\Pg\PH$ production
mechanism and select the DF $\Pe\Pgm$ final state. The main
background processes are the nonresonant $\PW\PW$, top quark (both single and pair production),
DY to $\tau$ lepton pairs, and \wjets when a jet is misidentified as a lepton. Smaller background contributions come from $\PW\PZ$, $\PZ\PZ$, V$\Pgg$, V$\Pgg^*$, and triboson
production. The $\PW\PW$ background process can be distinguished from the signal by the different kinematic
properties of the lepton system, since it is dominated by the on-shell $\PW$ boson pairs that do
not arise from a scalar resonance decay.
The top quark background process is
diluted by defining different categories that depend on the number of jets in
the event, and reduced by vetoing any $\PQb$-tagged jet with $\pt>20\GeV$.

The \wjets contribution (also referred to as nonprompt lepton background), where one jet mimics the
signature of an isolated prompt lepton, is an important background process especially
in the 0- and 1-jet $\Pg\Pg\PH$-tagged DF categories. This background is reduced
by taking advantage of the charge symmetry of the signal,
and the charge asymmetry of the \wjets process, in which the production of $\PWp$ is favored over $\PWm$.
Also, the fact
that the probabilities for a jet to mimic an electron or a muon are different,
and the fact that the misidentification rate is larger for lower-\pt leptons, are exploited.
Following these physics motivations the 0- and 1-jet $\Pg\Pg\PH$-tagged DF categories
are further split into four categories according to the lepton flavor, charge
and \pt ordering:
$\Pep\Pgmm$, $\Pem\Pgmp$, $\Pgmp\Pem$, and $\Pgmm\Pep$, where the
first lepton is the one with the higher \pt. In addition, the four
categories are divided according to whether the subleading lepton \pt ($\pt{}_2$) is
above or below $20\GeV$. This eight-fold partitioning of the 0-
and 1-jet $\Pg\Pg\PH$-tagged categories provides an improvement in terms of
the expected significance of about 15\% with respect to the inclusive 0- and 1-jet categories.

To suppress background processes with three or more leptons in the final state,
no additional identified and isolated
leptons with $\pt > 10\GeV$ are allowed in the events for the dilepton categories.
The dilepton invariant mass (\mll) is required to be higher than 12\GeV, to reject low-mass resonances and background that comes from events with multiple jets that all arise through the strong interaction (referred to the multijet background).
To suppress the background arising from DY events decaying to a $\tau$ lepton pair, which subsequently
decays to the $\Pe\Pgm$ final state, and to suppress processes without genuine missing transverse momentum, a minimum \ptmiss of 20\GeV
is required.
In the two-lepton categories, the DY background is further reduced by requiring the dilepton \pt (\ptll)
to be higher than 30\GeV, as on average $\Pe\Pgm$ lepton pairs from $\PZ\to\TT$ decays have
lower \pt than the ones from $\PH\to\PW\PW$ decays. These selection criteria
also reduce contributions from
$\PH\to\PW\PW\to\tau\nu\tau\nu$ and $\PH\to\TT$.
Finally, to further suppress contributions from $\PZ\to\TT$
and \wjets events, where the subleading lepton does not arise from a $\PW$ boson decay, the transverse mass built with \ptvecmiss and the subleading lepton, defined as:
\begin{equation}
\mtwTwo = \sqrt{2 \pt{_{2}} \ptmiss [1-\cos\delphiltwomet]},
\end{equation}
\noindent is required to be greater than 30\GeV. Here $\delphiltwomet$ is the azimuthal angle between the subleading lepton momentum and \ptvecmiss.

Although the invariant mass of the Higgs boson cannot
be reconstructed because of the undetected neutrinos, the expected kinematic properties of the Higgs boson
production and decay can be exploited.
The spin-0 nature of the SM Higgs boson results in the preferential emission of the two charged leptons in the same hemisphere.
Moreover, the invariant mass of the two leptons in the signal is
relatively small with respect to the one expected for a lepton pair arising from other
processes, such as nonresonant $\PW\PW$ and top quark production.
On the other hand, several of the smaller remaining background processes,
such as nonprompt leptons, DY$\to\TT$, and V$\Pgg$ populate the
same \mll phase space as the Higgs boson signal. These can be partially
disentangled from the signal by reconstructing the Higgs boson transverse mass as:
\begin{equation}
\mt = \sqrt{2 \ptll \ptmiss [1-\cos\delphillmet]},
\end{equation}
\noindent where $\delphillmet$ is the azimuthal angle between the dilepton
momentum and \ptvecmiss. These additional background processes populate
different regions of the two-dimensional plane in \mll and \mt.
A shape analysis based on a two-dimensional binned template fit of \mll versus \mt is
performed to extract the Higgs boson signal in the DF $\Pg\Pg\PH$ categories.

\begin{table*}[htb!]
\topcaption{
    Analysis categorization and event requirements for the 0-, 1-, and 2-jet $\Pg\Pg\PH$-tagged categories in the DF dilepton final state. The phase spaces defined by the 0-, 1-, and 2-jet $\Pg\Pg\PH$-tagged requirements correspond to the events shown in Figs.~\ref{fig:ggHOF_0j}, \ref{fig:ggHOF_1j}, and \ref{fig:ggHOF_2j}, respectively.
}\label{Tab:AnalysisStrategy:selections:gghof}
\centering
\cmsTable{
\begin{tabular}{ccc}
\hline
Category & Subcategory & Requirements   \\
\hline
\multirow{4}{*}{Preselection}             &   \multirow{4}{*}{\NA}  & $\mll > 12\GeV$, $\pt{}_1 > 25\GeV$, $\pt{}_2 > 13~(10)\GeV$ for $\Pe$ ($\Pgm$),  \\
                                          &                         & $\ptmiss > 20\GeV$, $\ptll > 30\GeV$  \\
					  &			    & no additional leptons with $\pt>10\GeV$ \\
                                          &                         & electron and muon with opposite charges    \\[\extraTabSkip]
\multirow{8}{*}{0-jet $\Pg\Pg\PH$-tagged}  & \multirow{4}{*}{ $\left.\begin{tabular}{c} $\Pep\Pgmm$\\$\Pem\Pgmp$\\$\Pgmp\Pem$\\$\Pgmm\Pep$ \end{tabular}\right\} \pt{_{2}} > 20\GeV$} & $\mt > 60\GeV$, $\mtwTwo> 30\GeV$    \\
                                   &  & subleading lepton $\pt > 20\GeV$  \\
                                   &  & no jets with $\pt > 30\GeV$ \\
                                   &  & no $\PQb$-tagged  jets with \pt between 20 and 30\GeV  \\[\extraTabSkip]
                                   & \multirow{4}{*}{ $\left.\begin{tabular}{c} $\Pep\Pgmm$\\$\Pem\Pgmp$\\$\Pgmp\Pem$\\$\Pgmm\Pep$ \end{tabular}\right\} \pt{_{2}} < 20\GeV$} & $\mt > 60\GeV$, $\mtwTwo > 30\GeV$    \\
                                   &  & subleading lepton $\pt < 20\GeV$  \\
                                   &  & no jets with $\pt > 30\GeV$ \\
                                   &  & no $\PQb$-tagged  jets with \pt between 20 and 30\GeV  \\[\extraTabSkip]
\multirow{8}{*}{1-jet $\Pg\Pg\PH$-tagged}   & \multirow{4}{*}{ $\left.\begin{tabular}{c} $\Pep\Pgmm$\\$\Pem\Pgmp$\\$\Pgmp\Pem$\\$\Pgmm\Pep$ \end{tabular}\right\} \pt{_{2}} > 20\GeV$} & $\mt > 60\GeV$, $\mtwTwo > 30\GeV$    \\
                                    &  & subleading lepton $\pt > 20\GeV$   \\
                                    &  & exactly one jet with $\pt > 30\GeV$ \\
                                    &  & no $\PQb$-tagged  jets with $\pt > 20\GeV$  \\[\extraTabSkip]
                                    & \multirow{4}{*}{ $\left.\begin{tabular}{c} $\Pep\Pgmm$\\$\Pem\Pgmp$\\$\Pgmp\Pem$\\$\Pgmm\Pep$ \end{tabular}\right\} \pt{_{2}} < 20\GeV$} & $\mt > 60\GeV$, $\mtwTwo > 30\GeV$    \\
                                    &  & subleading lepton $\pt < 20\GeV$   \\
                                    &  & exactly one jet with $\pt > 30\GeV$ \\
                                    &  & no $\PQb$-tagged  jets with $\pt > 20\GeV$  \\[\extraTabSkip]
\multirow{4}{*}{2-jet $\Pg\Pg\PH$-tagged}  & \multirow{4}{*}{$\Pe\Pgm$}     & at least two jets with $\pt > 30\GeV$    \\
                                   &                             & $\mtwTwo > 30\GeV$ and $\mt > 60\GeV$    \\
                                   &                             & no $\PQb$-tagged  jets with $\pt > 20\GeV$  \\
                                   &                             & $\mjj < 65\GeV$ or $105 < \mjj < 400\GeV$  \\
\hline
\end{tabular}
}
\end{table*}

\begin{figure*}[!t]
\centering
\includegraphics[width=\cmsFigWidthAlt]{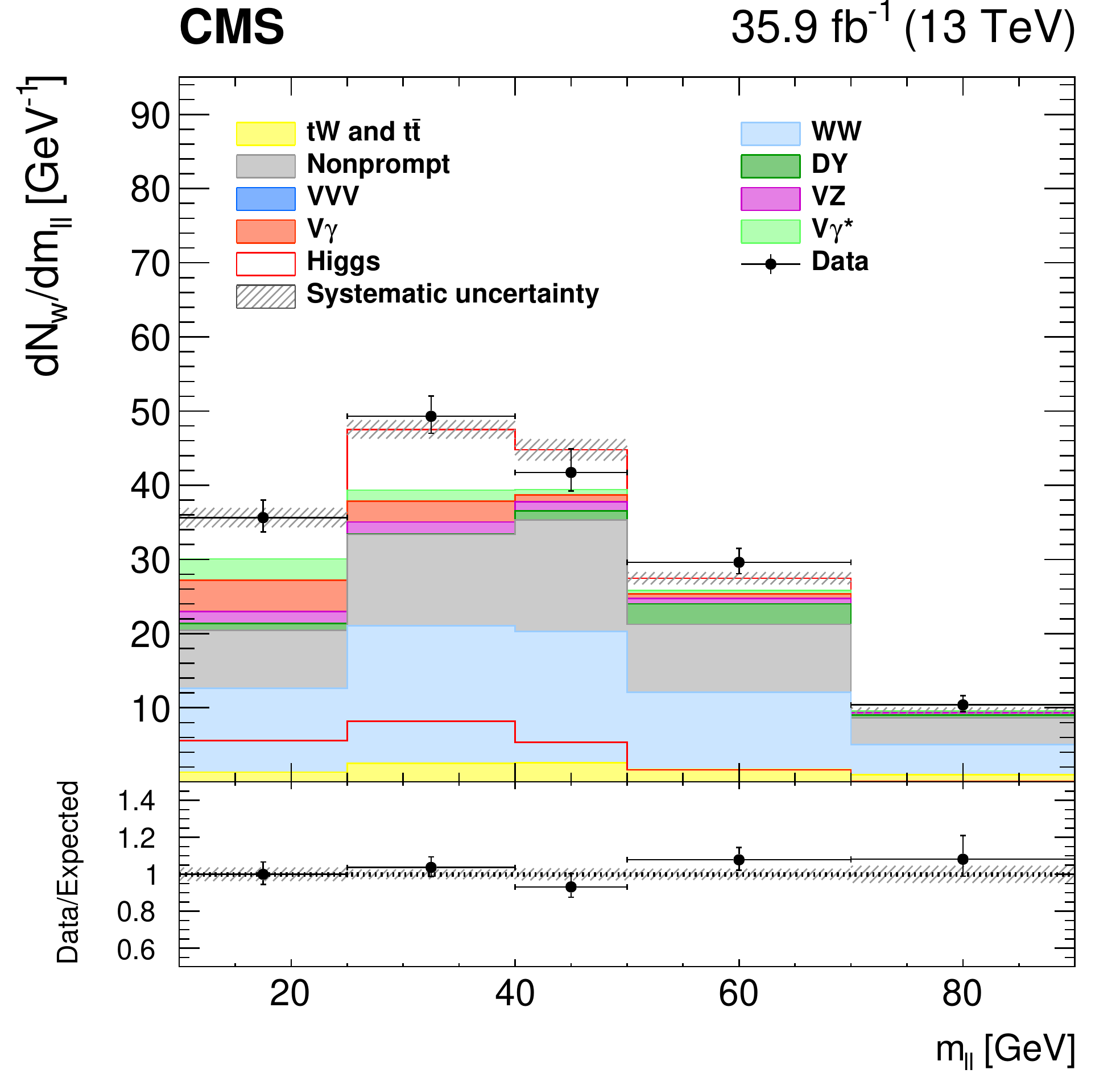}
\includegraphics[width=\cmsFigWidthAlt]{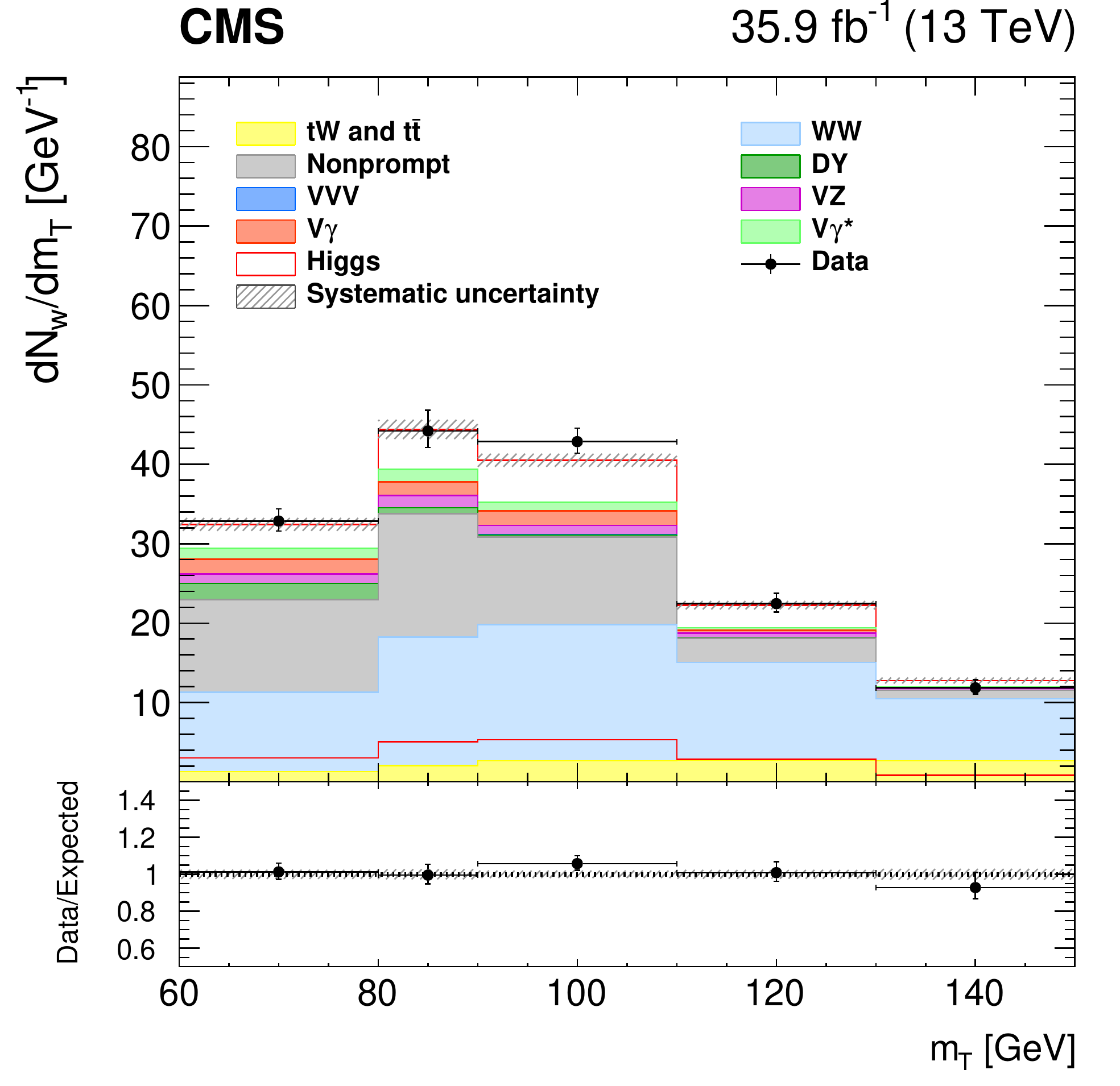}\\
\includegraphics[width=\cmsFigWidthAlt]{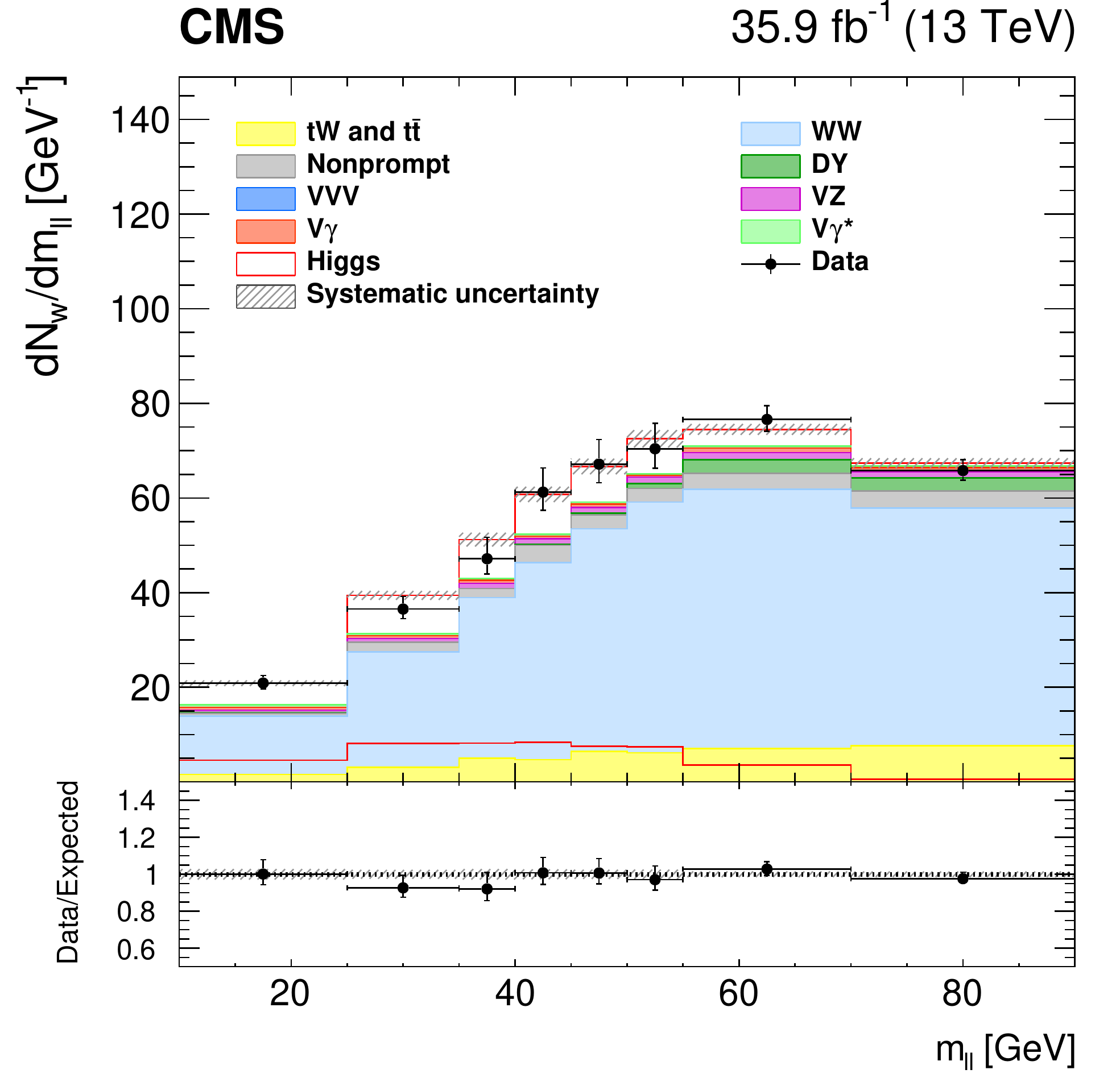}
\includegraphics[width=\cmsFigWidthAlt]{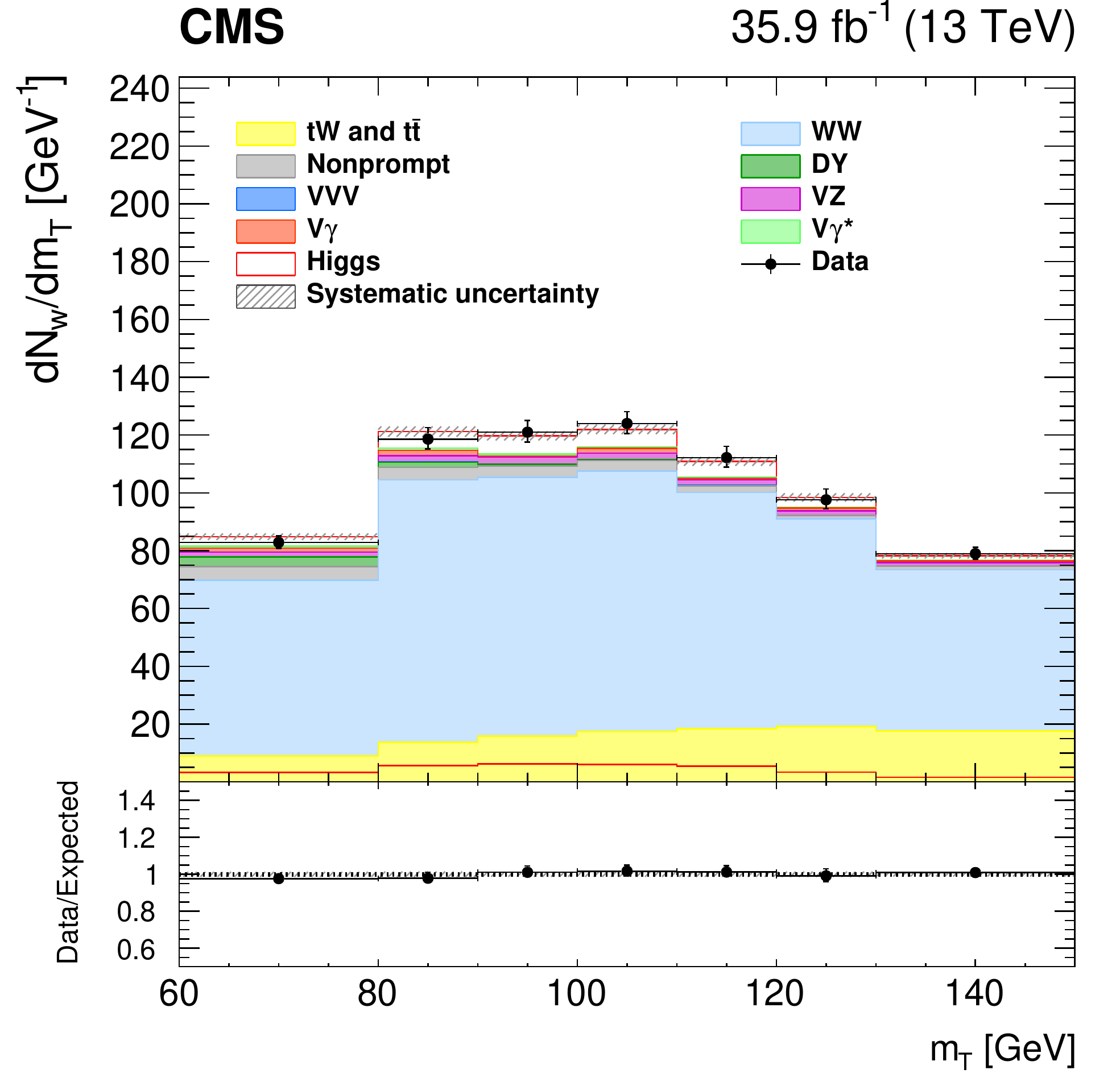}\\
\caption{
Postfit number of weighted events ($N_\mathrm{w}$) as a function of \mll and
\mt for DF events with 0 jets and $\pt{_{2}} < 20\GeV$ (upper row) or
$\pt{_{2}} > 20\GeV$ (lower row). The number of events is weighted according to the ${\mathrm{S}/(\mathrm{S}+\mathrm{B})}$ ratio in each bin of one of the two variables, integrating over the other one.
The various lepton flavor and charge
subcategories are also merged and weighted according to their ${\mathrm{S}/(\mathrm{S}+\mathrm{B})}$ value.
The contributions of the main background processes (stacked histograms) and the Higgs boson signal
(superimposed and stacked red histograms) remaining after all selection criteria are shown.
The dashed gray band accounts for all systematic uncertainties on the signal and background yields after the fit.
}
\label{fig:ggHOF_0j}
\end{figure*}

\begin{figure*}[!t]
\centering
\includegraphics[width=\cmsFigWidthAlt]{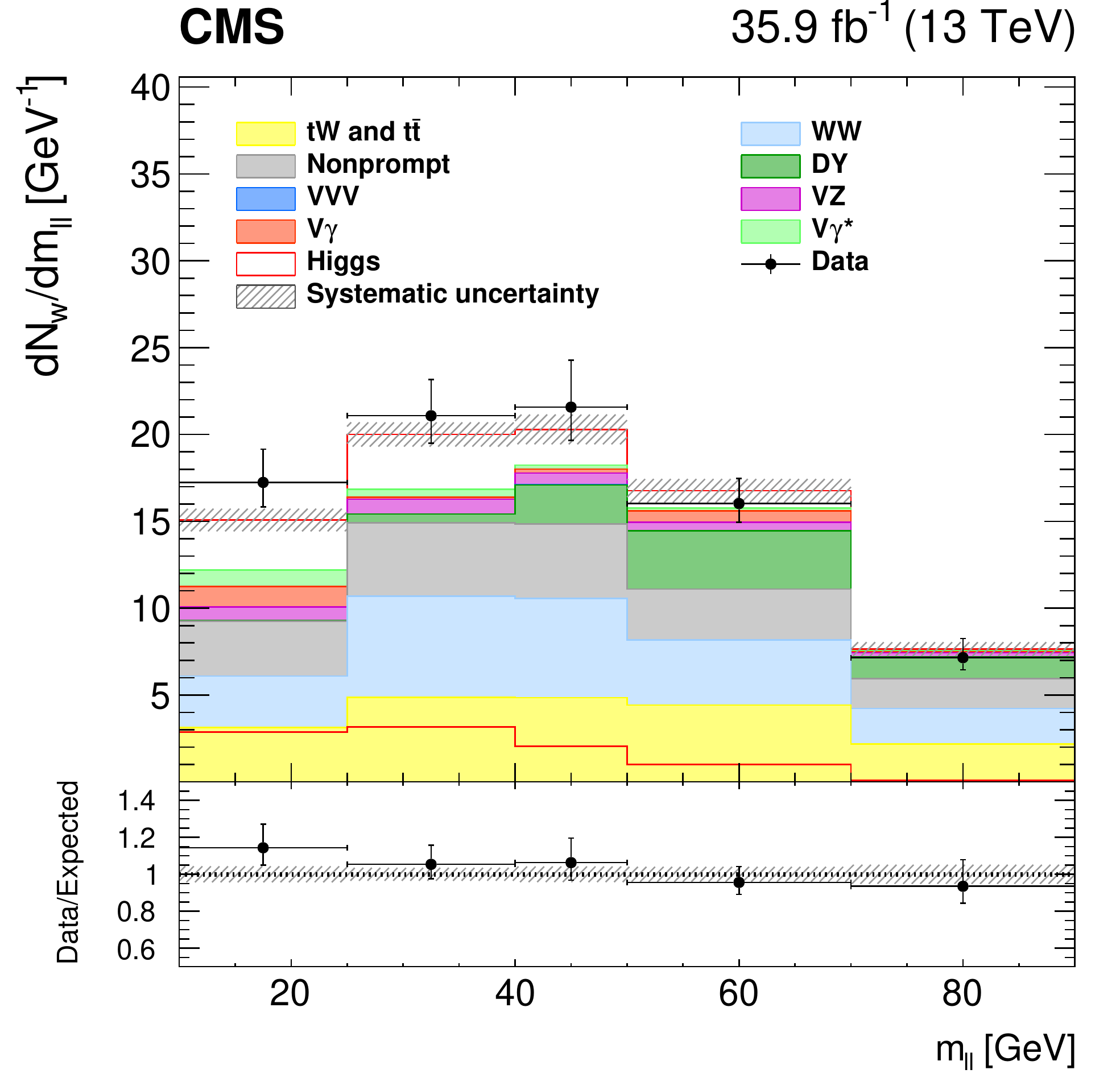}
\includegraphics[width=\cmsFigWidthAlt]{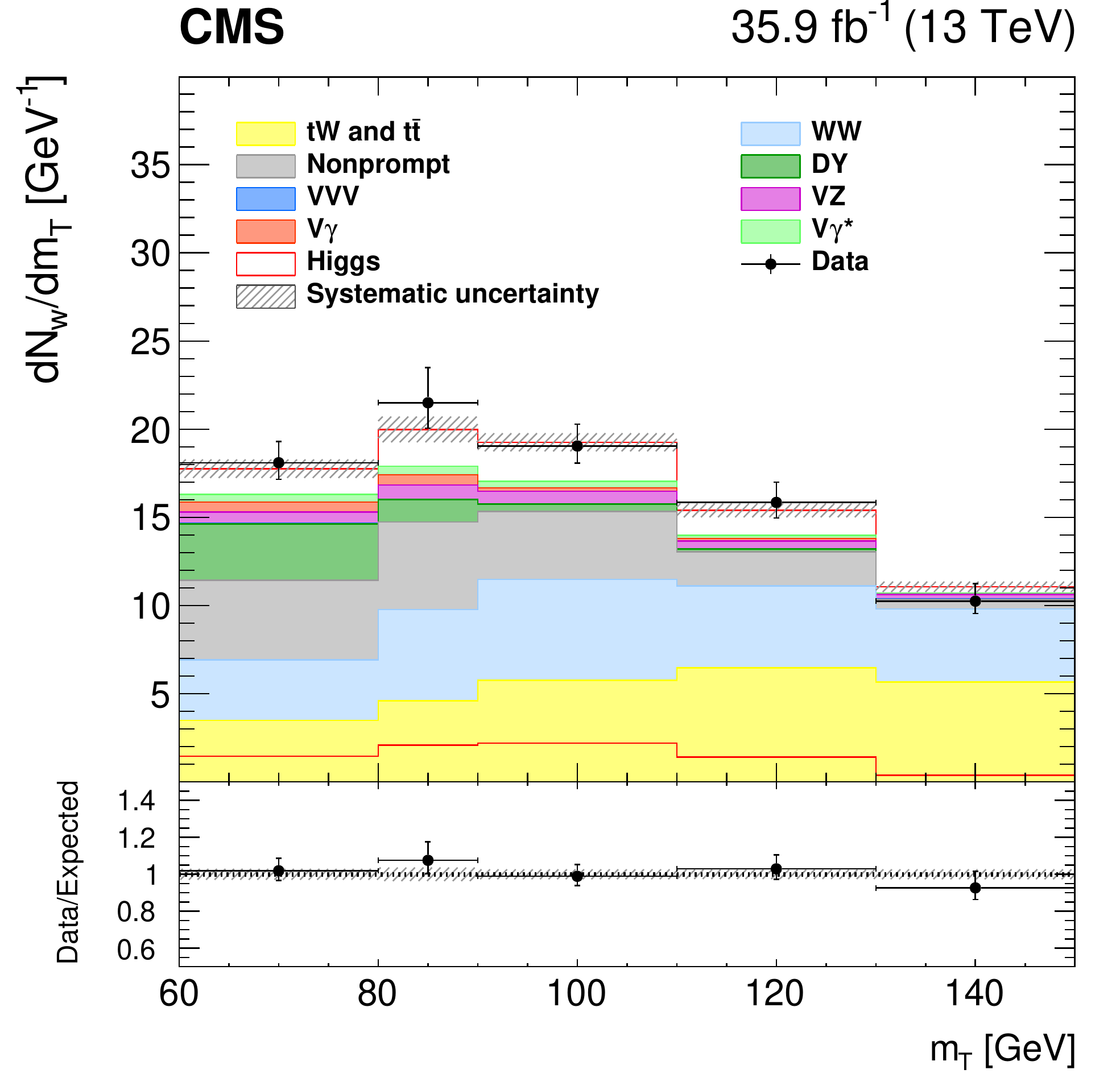}\\
\includegraphics[width=\cmsFigWidthAlt]{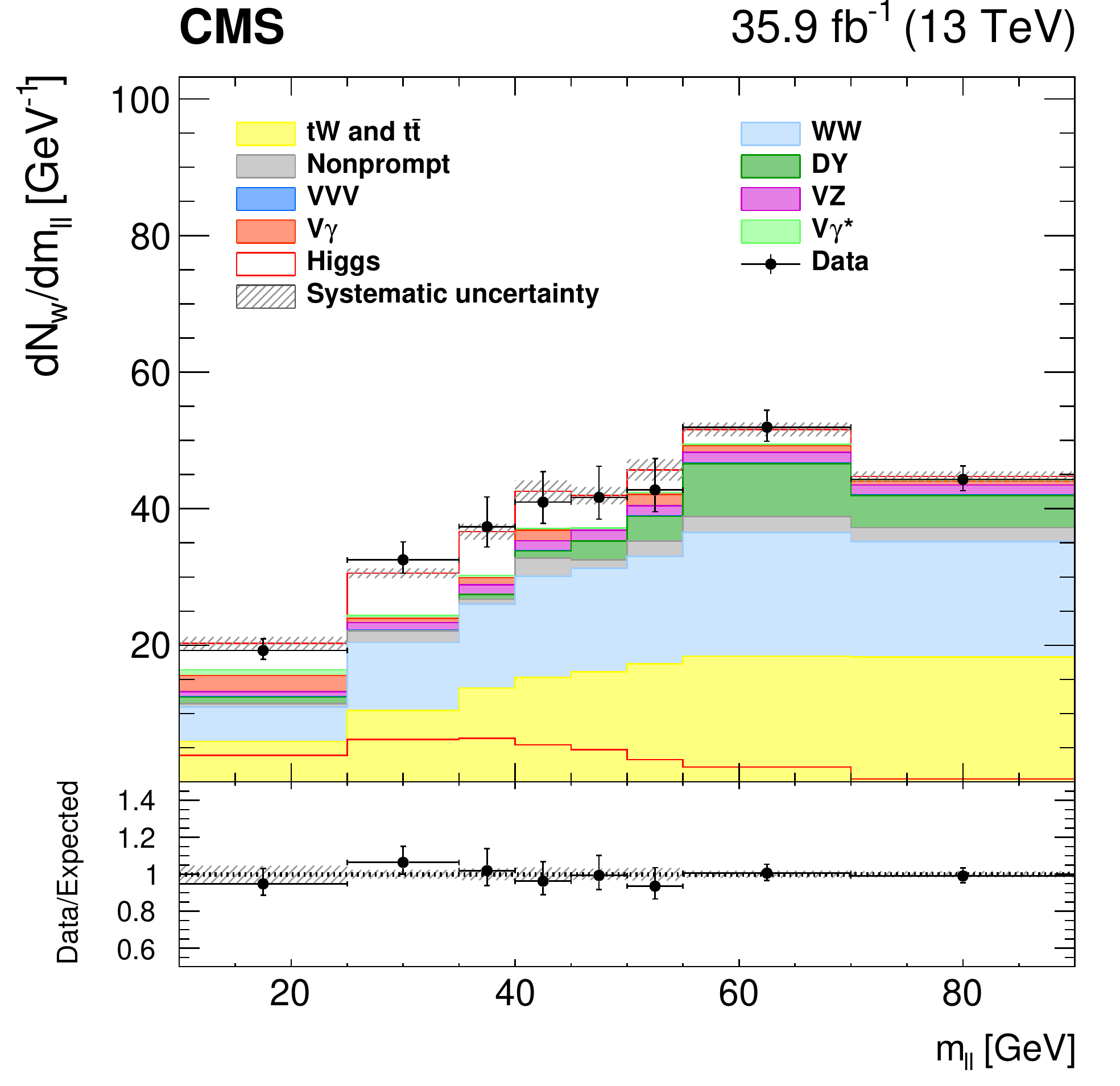}
\includegraphics[width=\cmsFigWidthAlt]{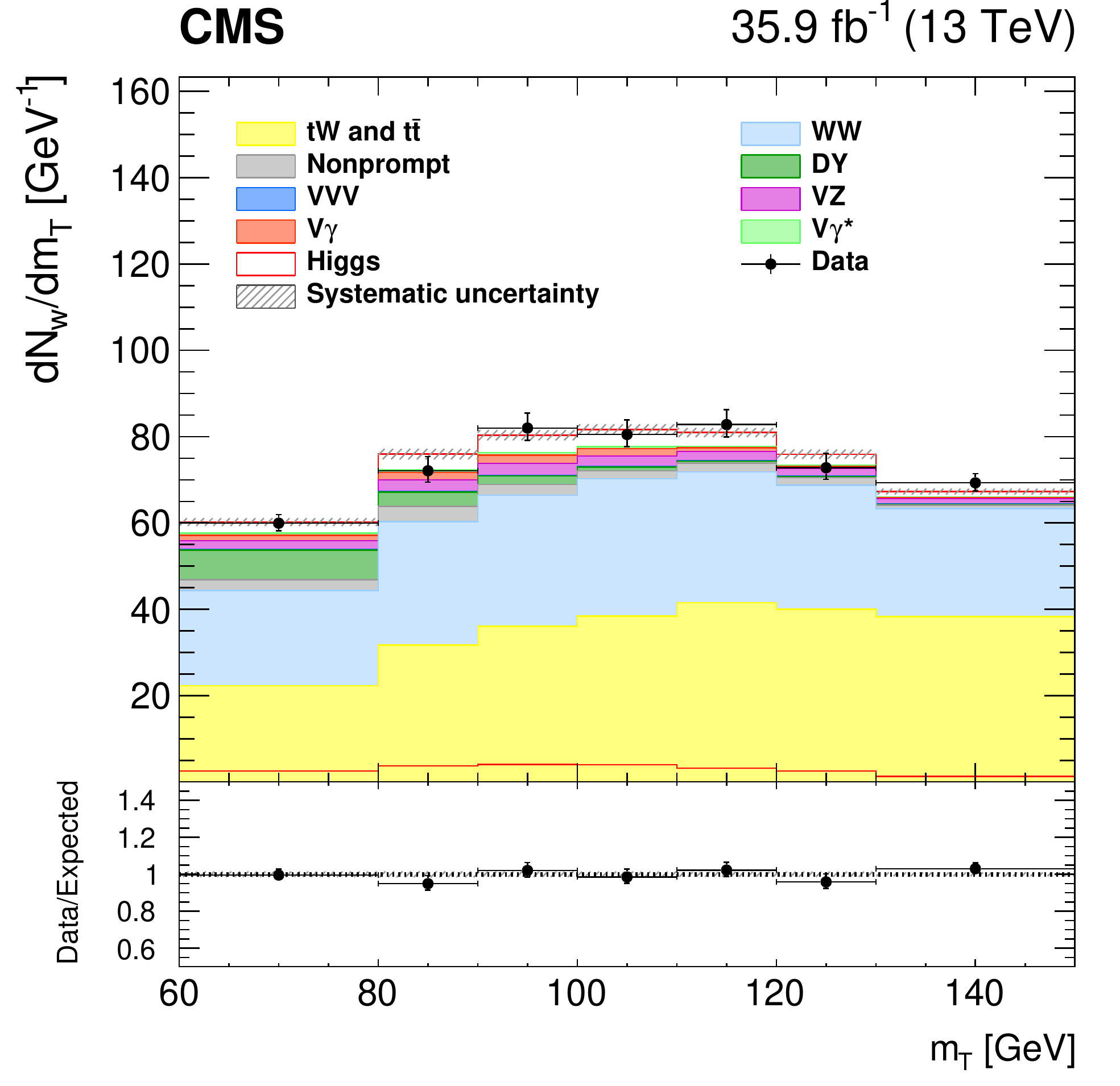}\\
\caption{
Same as previous figure, for DF events with one jet.
}
\label{fig:ggHOF_1j}
\end{figure*}

\begin{figure*}[!t]
\centering
\includegraphics[width=\cmsFigWidthAlt]{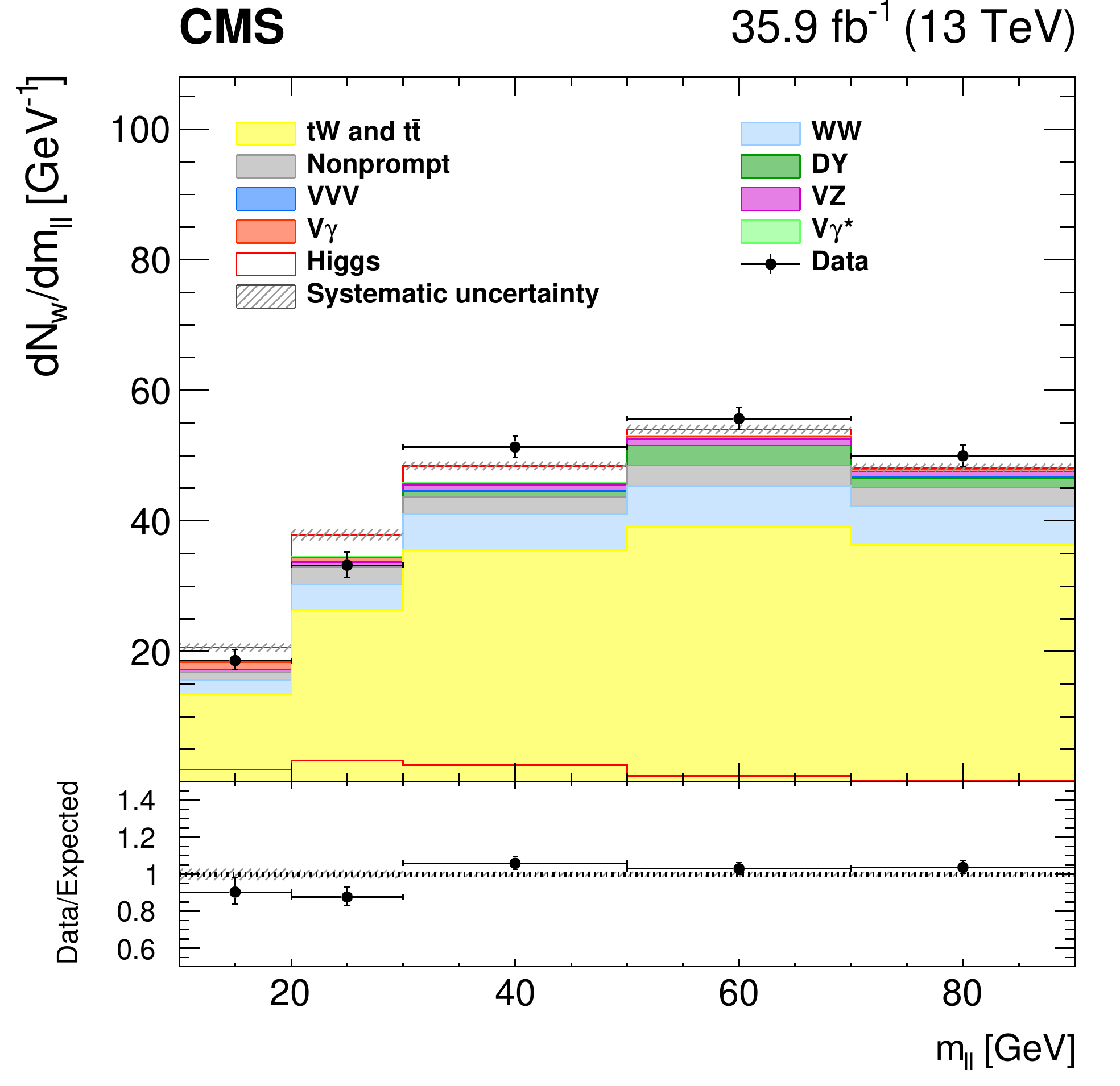}
\includegraphics[width=\cmsFigWidthAlt]{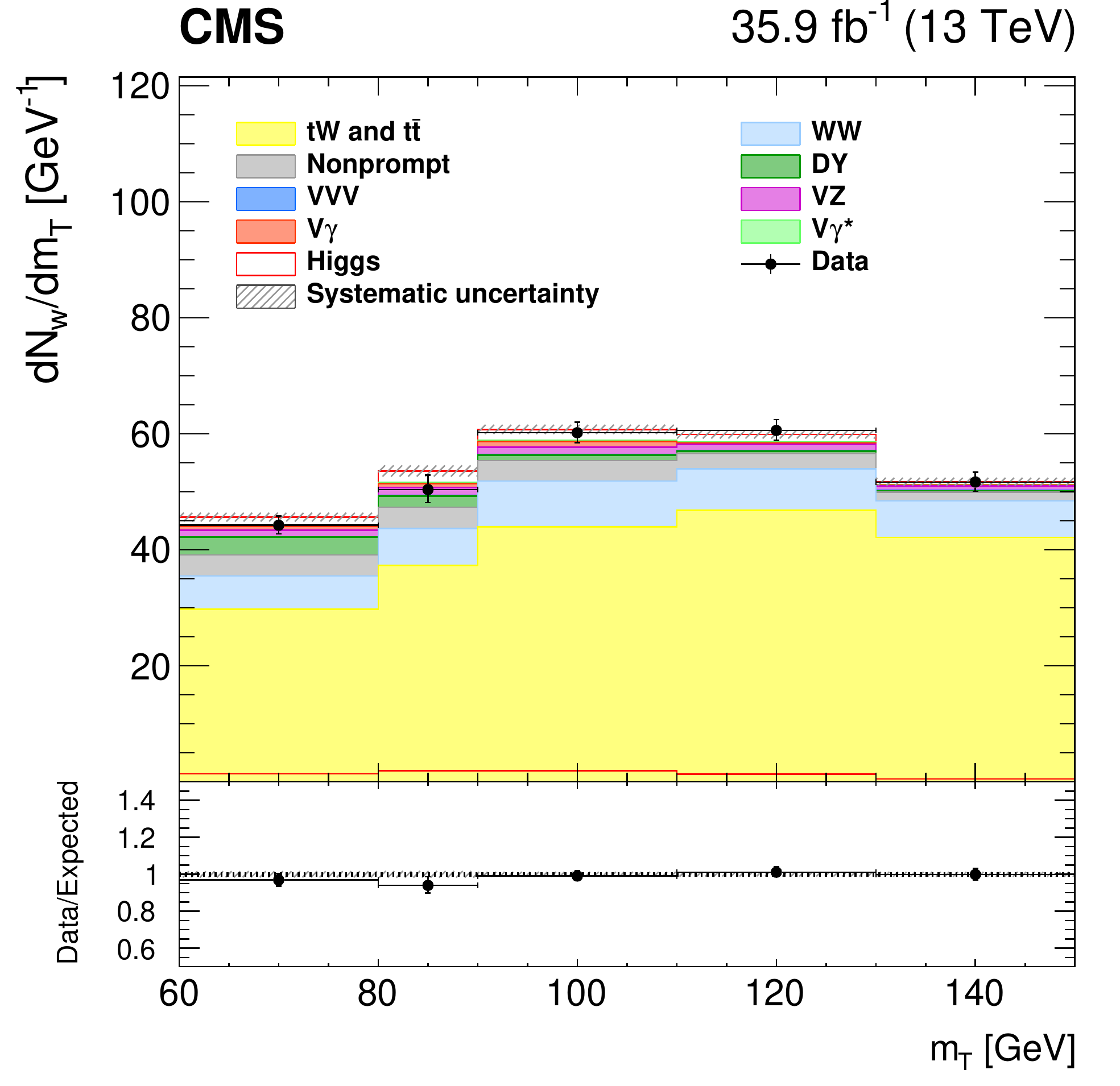}
\caption{
Postfit number of weighted events ($N_\mathrm{w}$) as a function of \mll and \mt for DF events with at least 2 jets. The number of events is weighted according to the ${\mathrm{S}/(\mathrm{S}+\mathrm{B})}$ ratio in each bin of one of the two variables, integrating over the other one.
}
\label{fig:ggHOF_2j}
\end{figure*}

The observed events
as a function of \mll and \mt are shown in Figs.~\ref{fig:ggHOF_0j}, \ref{fig:ggHOF_1j}, and \ref{fig:ggHOF_2j}, after the template fit to the (\mll,~\mt) distribution. The 0- and 1-jet categories are split into $\pt{_{2}} < 20$\GeV and $\pt{_{2}} > 20$\GeV subcategories, to show the
different purity of the two regions.
In these figures the postfit number of events is shown, \ie, each signal and background process is normalized to the result of a simultaneous fit to all categories, assuming that the relative proportions for the different Higgs boson production mechanisms are those predicted by the SM. The events in each bin of one of the two variables are obtained by integrating over the other, and weighted using the ratio of fitted signal $\mathrm{(S)}$ to the sum of signal and background $\mathrm{(S+B)}$.
${\mathrm{S}/(\mathrm{S}+\mathrm{B})}$ ratio in each \mt bin. This ratio is then used to perform a weighted sum of the \mll distributions in each \mt bin. A similar weighting procedure is applied when merging the distributions of a given variable in different categories.
The weighting procedure is used only for visualization purposes, and is not used for signal extraction.

The full list of DF $\Pg\Pg\PH$ categories and their selection requirements
is shown in Table~\ref{Tab:AnalysisStrategy:selections:gghof}.

\subsection{Different-flavor VBF category}\label{sec:AnalysisStrategyVBFOF}

\begin{figure*}[!h]
\centering
\includegraphics[width=\cmsFigWidthAlt]{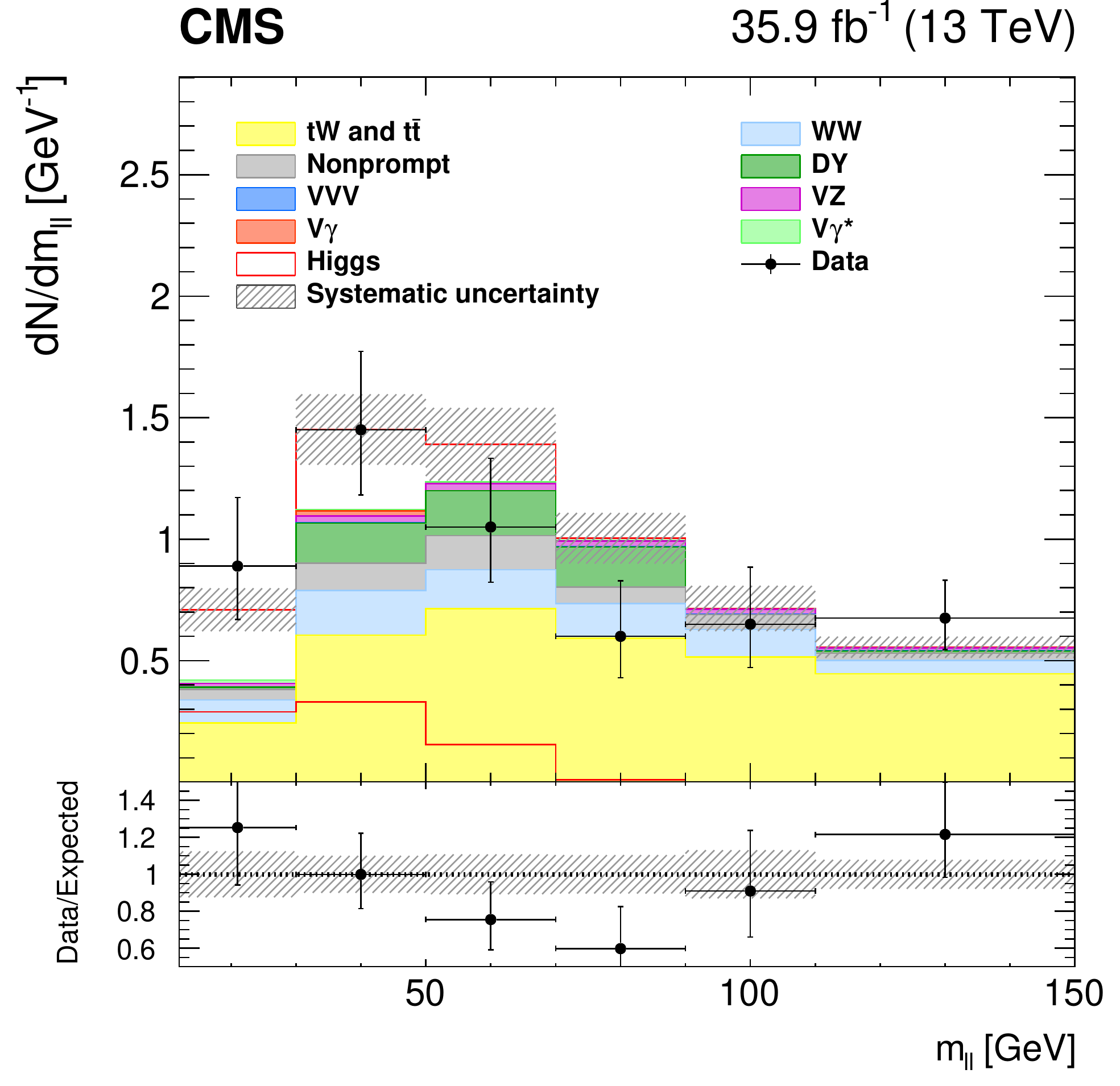}
\includegraphics[width=\cmsFigWidthAlt]{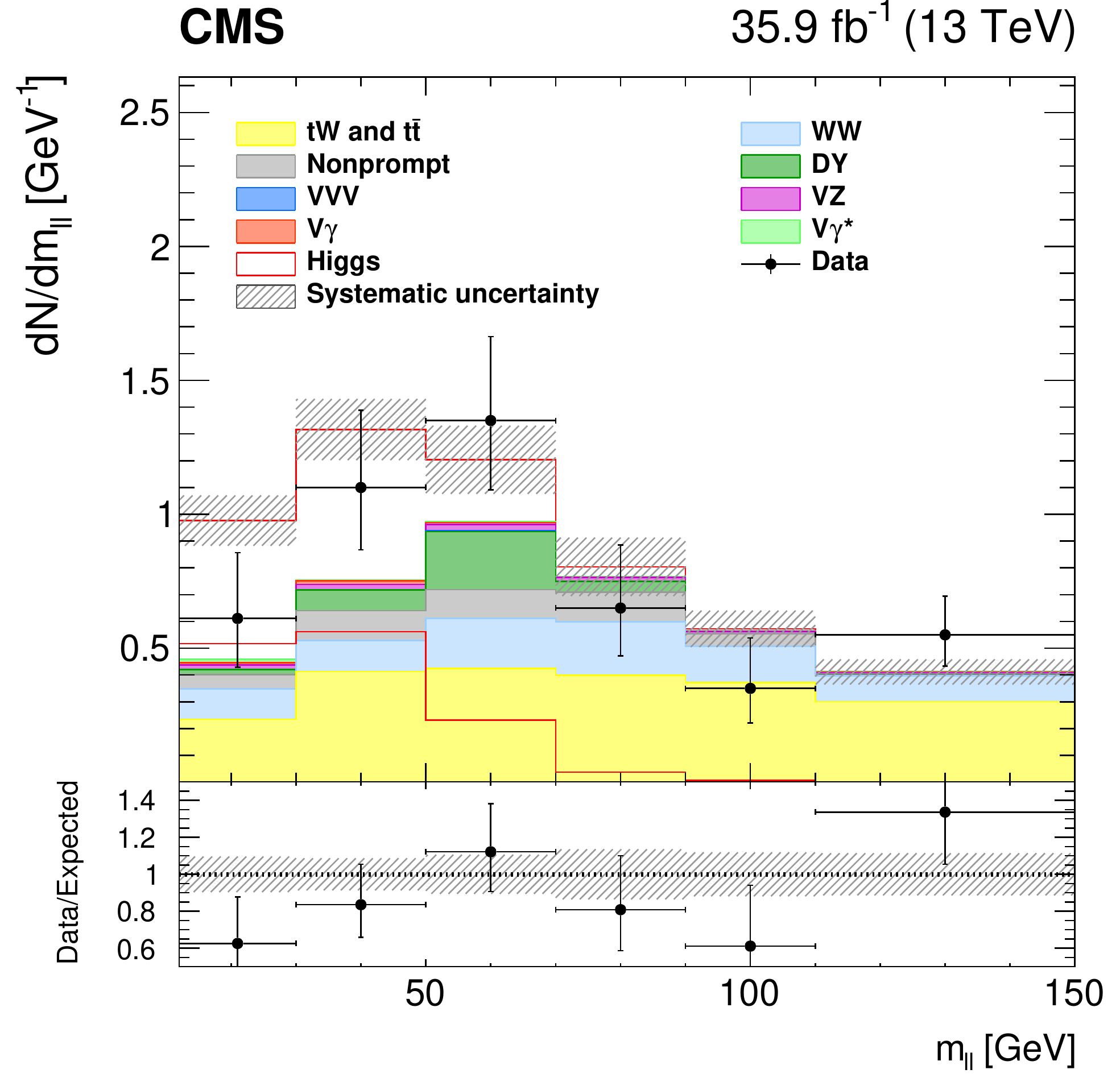}
\caption{
        Postfit number of events with VBF topology as a function of \mll, for
        $400 < \mjj < 700\GeV$ (left) and $\mjj > 700\GeV$ (right).
        }  \label{fig:vbf}
\end{figure*}

The VBF process is the second largest Higgs boson production mechanism at the
LHC.
This mode involves the production of a Higgs boson in association with two jets with large rapidity separations.
After the common preselection,
the VBF analysis requires events with exactly two jets with $\pt > 30\GeV$,
a pseudorapidity separation ($\abs{\detajj}$) between the two jets
larger than 3.5,
and an invariant mass ($\mjj$) greater than 400\GeV.
The rejection of events with more than two jets reduces the \ttbar background contribution without affecting the signal efficiency, thus improving the signal sensitivity.
The VBF analysis is based on the shape of the \mll distribution,
 and is split into two signal regions, one with ${400 < \mjj < 700\GeV}$ and
 the other with $\mjj > 700\GeV$, to profit from the higher purity of the
 $\mjj > 700\GeV$ region.
The post-fit signal and background events as functions of \mll are shown in Fig.~\ref{fig:vbf}, for the two \mjj regions separately.
The list of event requirements applied in this category is presented in
 Table~\ref{Tab:AnalysisStrategy:selections:vbfof}.

\begin{table*}[htbp]
\small
\topcaption{
    Analysis categorization and event requirements for the 2-jet VBF-tagged category, in the DF dilepton final state. The phase spaces defined by the 2-jet VBF-tagged requirements correspond to the events shown in Fig.~\ref{fig:vbf}.
}\label{Tab:AnalysisStrategy:selections:vbfof}
\centering
\begin{tabular}{ccc}
\hline
Category & Subcategory & Requirements   \\
\hline

\multirow{4}{*}{Preselection}  &   \multirow{4}{*}{\NA}    & $\mll >12\GeV$, $\pt{}_1 > 25\GeV$, $\pt{}_2 > 13~(10)\GeV$ for $\Pe$($\Pgm$),      \\
                                          &                         & $\ptmiss > 20\GeV$, $\ptll > 30\GeV$ \\
					  &			    & no additional leptons with $\pt>10\GeV$ \\
                                          &                         & electron and muon with opposite charges    \\[\extraTabSkip]
\multirow{10}{*}{2-jet VBF-tagged} & \multirow{5}{*}{$\Pe\Pgm$ low \mjj}     & exactly two jets with $\pt > 30\GeV$    \\
                        &                             & $60 < \mt < 125\GeV$    \\
                        &                             & leptons $\eta$ between the two leading jets  \\
                        &                             & $400 < \mjj < 700\GeV$ and $\abs{\detajj} > 3.5$  \\
                        &                             & no $\PQb$-tagged  jets with $\pt > 20\GeV$  \\[\extraTabSkip]
                        & \multirow{5}{*}{$\Pe\Pgm$ high \mjj}     & exactly two jets with $\pt > 30\GeV$    \\
                        &                             & $60 < \mt < 125\GeV$    \\
                        &                             & leptons $\eta$ between the two leading jets  \\
                        &                             & $\mjj > 700\GeV$ and $\abs{\detajj} > 3.5$  \\
                        &                             & no $\PQb$-tagged  jets with $\pt > 20\GeV$  \\
\hline
\end{tabular}
\end{table*}

\subsection{Different-flavor \texorpdfstring{V$\PH$}{VH} with two jets category}\label{sec:AnalysisStrategyVH2jOF}

The V$\PH$ process involves
the production of a Higgs boson in association with a $\PW$ or $\PZ$ boson.
The 2-jet V$\PH$-tagged category targets final states where one vector boson ($\PW$ or $\PZ$) decays into two resolved jets.
This category with hadronically decaying vector bosons is affected by large backgrounds compared to the leptonic decays, but profits from a higher branching fraction.
The 2-jet V$\PH$-tagged analysis reverses the pseudorapidity separation requirement of the VBF selection ($\abs{\Delta\eta} < 3.5$) and
requires \mjj to be between 65 and 105\GeV. In addition, the
two leading jets are required to be central ($\abs{\eta}< 2.5$) to profit
from more stringent $\PQb$ jet veto requirements, given that $\PQb$ tagging can only be performed for central jets.
A cut on $\Delta R_{\ell\ell} < 2$ is applied to suppress \ttbar background,
taking advantage of the spin-0 nature of the Higgs boson that results in leptons
being preferentially emitted in nearby directions. This kinematic property is
further enhanced in this category due to the boost of the Higgs boson
recoiling against the associated vector boson.

\begin{figure}[!b]
\centering
\includegraphics[width=\cmsFigWidthAlt]{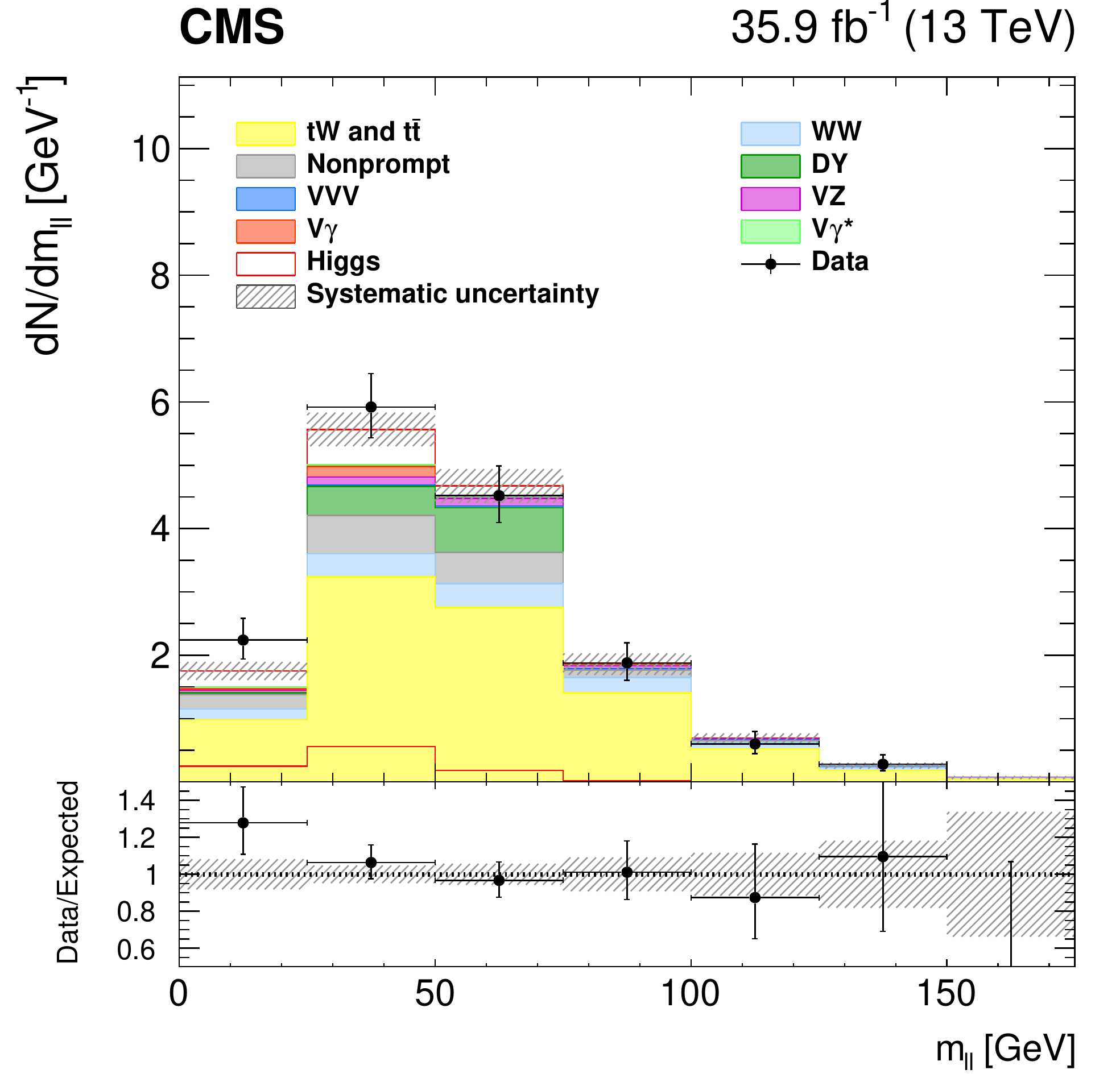}
\caption{
       Postfit number of events as a function of \mll for DF events in the
       2-jets V$\PH$-tagged category.
        }  \label{fig:vh2j}
\end{figure}

The analysis is based on the shape of the \mll discriminant distribution,
presented in Fig.~\ref{fig:vh2j}. The list of event requirements applied
is presented in Table~\ref{Tab:AnalysisStrategy:selections:vh2jof}.

\begin{table*}[htbp]
\small
\topcaption{
    Analysis categorization and event requirements for the 2-jet V$\PH$-tagged category, in the DF dilepton final state. The phase space defined by the 2-jet V$\PH$-tagged requirements corresponds to the events shown in Fig.~\ref{fig:vh2j}.
    }\label{Tab:AnalysisStrategy:selections:vh2jof}
\centering
\begin{tabular}{ccc}
\hline
Category & Subcategory & Requirements   \\
\hline

\multirow{4}{*}{Preselection}  &   \multirow{4}{*}{\NA}    & $\mll >12\GeV$, $\pt{}_1 > 25\GeV$, $\pt{}_2 > 13~(10)\GeV$ for $\Pe$ ($\Pgm$)      \\
                                          &                         & $\ptmiss > 20\GeV$, $\ptll > 30\GeV$  \\
					  &			    & no additional leptons with $\pt>10\GeV$ \\
                                          &                         & electron and muon with opposite charges    \\[\extraTabSkip]
\multirow{5}{*}{2-jet V$\PH$-tagged}  & \multirow{5}{*}{$\Pe\Pgm$}     & at least two jets with $\pt > 30\GeV$     \\
                        &                             & two leading jets with $\abs{\eta} < 2.5$    \\
                        &                             & $60 < \mt < 125\GeV$  and $\Delta R_{\ell\ell} < 2$  \\
                        &                             & no $\PQb$-tagged  jets with $\pt > 20\GeV$  \\
                        &                             & $65 < \mjj < 105\GeV$ and $\abs{\detajj} < 3.5$ \\

\hline
\end{tabular}
\end{table*}

\begin{table*}[htbp]
\small
\centering
\topcaption{
    Analysis categorization and selections for the 0- and 1- jet $\Pg\Pg\PH$-tagged categories in the SF dilepton final state.
    }\label{Tab:AnalysisStrategy:selections:SF012}
\cmsTable{
\begin{tabular}{ccc}
\hline
Category & Subcategory & Requirements   \\
\hline

\multirow{4}{*}{Preselection}      &   \multirow{4}{*}{\NA}               & $\mll > 12\GeV$, $\pt{}_1 > 25~(20)\GeV$ for $\Pe$ ($\Pgm$), $\pt{}_2 > 13~(10)\GeV$ for $\Pe$ ($\Pgm$),  \\
                                   &                                    & track $\ptmiss > 20\GeV$, $\ptll > 30\GeV$ \\
			           &	 				    & no additional leptons with $\pt>10\GeV$ \\
                                   &                                    & two electrons or two muons with opposite charges    \\[\extraTabSkip]
\multirow{8}{*}{0-jet $\Pg\Pg\PH$-tagged}  &  						& DYMVA\,$>0.991$, $\mll<55\GeV$, $\mt>50\GeV$,   \\
                                   &      $\EE$ $\pt{_{2}} < 20\GeV$	        & $\pt{_{2}} < 20\GeV$, $\Delta\phi{_{\ell\ell}} < 1.7$    \\
                   		     &    $\MM$ $\pt{_{2}} < 20\GeV$               			& no jets with $\pt > 30\GeV$ \\
                        	     &                  			& no $\PQb$-tagged  jets with $\pt > 20\GeV$  \\[\extraTabSkip]
                                   & 	      & DYMVA\,$>0.991$, $\mll<55\GeV$, $\mt>50\GeV$,  \\
                                   &  $\EE$ $\pt{_{2}} > 20\GeV$ 	& $20\GeV<\pt{_{2}}<50\GeV$, $\Delta\phi{_{\ell\ell}} < 1.7$    \\
		                       & $\MM$ $\pt{_{2}} > 20\GeV$           	                  & no jets with $\pt > 30\GeV$ \\
                		           &		                        & no $\PQb$-tagged  jets with $\pt > 20\GeV$  \\[\extraTabSkip]
\multirow{4}{*}{1-jet $\Pg\Pg\PH$-tagged}  &  				            & DYMVA\,$>0.95$, $\mll<57\GeV$, $50 < \mt < 155\GeV$,    \\
                                   & 	$\EE$			      & $\pt{_{1}}<50\GeV$, $\Delta\phi{_{\ell\ell}} < 1.75$    \\
                       		     &	$\MM$ 		                  & exactly one jet with $\pt > 30\GeV$ \\
                                   &			                  & no $\PQb$-tagged  jets with $\pt > 20\GeV$  \\
\hline
\end{tabular}
}
\end{table*}

\begin{table*}[!ht]
\small
\topcaption{
    Analysis categorization and event requirements for the $\PW\PH$-tagged
    category, in the three-lepton final state. Here, \mllminlll is the minimum
    \mll between the oppositely charged leptons. For the $\PZ$ boson veto, the
    opposite-sign same-flavor pair with the \mll closest to the $\PZ$ boson mass is considered. Events that fulfill the three-lepton $\PW\PH$-tagged requirements correspond to the signal phase space shown in Fig.~\ref{fig:wh3l}.
    }\label{Tab:AnalysisStrategy:selections:wh3l}
\centering
\begin{tabular}{ccc}
\hline
Category & Subcategory & Requirements   \\
\hline

\multirow{3}{*}{Preselection}             &   \multirow{3}{*}{\NA}    & $\pt{}_1 > 25\GeV$, $\pt{}_2 >20\GeV$, $\pt{}_3 > 15\GeV$     \\
                                          &                         & no additional leptons with $\pt>10\GeV$ \\
                                          &                         & $\mllminlll >12\GeV$ , total lepton charge sum $\pm 1$\\[\extraTabSkip]
\multirow{8}{*}{3-lepton $\PW\PH$-tagged} & \multirow{5}{*}{OSSF}       & no jets with $\pt > 30\GeV$    \\
	                                   &                             & no $\PQb$-tagged jets with $\pt > 20\GeV$  \\
               		                 &                             & $\ptmiss > 50\GeV$, $\mllminlll < 100\GeV$  \\
                            		     &                             & $\PZ$ boson veto: $\abs{\mll - m_{\PZ}} > 25\GeV$ \\
		                             &                             & $\Delta\phi(\ell\ell\ell,\ptvecmiss)> 2.2$ \\[\extraTabSkip]
                          		     & \multirow{3}{*}{SSSF}       & no jets with $\pt > 30\GeV$      \\
		                             &                             & no $\PQb$-tagged jets with $\pt > 20\GeV$ \\
                		                 &                             & $\Delta\phi(\ell\ell\ell,\ptvecmiss) > 2.5$ \\
\hline
 \end{tabular}
\end{table*}

\begin{table*}[htbp]
\small
\topcaption{
    Analysis categorization and event requirements for the $\PZ\PH$-tagged category, in the four-lepton final state. Here, $\mathrm{X}$ is defined as the remaining lepton pair after the $\PZ$ boson candidate is chosen.
The component leptons of $\mathrm{X}$ can be either same-flavor (XSF) or different-flavor (XDF).
    }
\label{Tab:AnalysisStrategy:selections:zh4l}
\centering
\begin{tabular}{ccc}
\hline
Category & Subcategory & Requirements   \\
\hline

\multirow{8}{*}{Preselection}             &   \multirow{8}{*}{\NA}    &  {four tight and isolated leptons, with zero total charge} \\
                                          &                         &  {$\pt > 25\GeV$ for the leading lepton} \\
                                          &                         &  {$\pt > 15\GeV$ for the second leading lepton} \\
                                          &                         &  {$\pt > 10\GeV$ for the remaining two leptons} \\
                                          &                         &  no additional leptons with $\pt>10\GeV$ \\
                                          &                         &  {$\PZ$ dilepton mass $>$4\GeV} \\
                                          &                         &  {$\mathrm{X}$ dilepton mass $>$4\GeV} \\
                                          &                         &  {no $\PQb$-tagged jets with $\pt > 20\GeV$} \\[\extraTabSkip]
\multirow{7}{*}{4-lepton $\PZ\PH$-tagged}   & \multirow{4}{*}{XSF}    &  { $\abs{\mll-m_{\PZ}}<15\GeV$}   \\
                       		            &                         &  {$10 < m_{X} < 50\GeV$} \\
                       		            &                         &  {$35 < \ptmiss < 100\GeV$}  \\
                                          &                         &  {four-lepton invariant mass $>$140\GeV} \\[\extraTabSkip]
                        	            & \multirow{3}{*}{XDF}    & {$\abs{\mll-m_{\PZ}}<15\GeV$}    \\
	                                    &                         & {$10 < m_\mathrm{X} < 70\GeV$} \\
          		                        &                         & {$\ptmiss >20\GeV$} \\

\hline
 \end{tabular}
\end{table*}

\subsection{Same-flavor \texorpdfstring{$\Pg\Pg\PH$}{ggH} categories}\label{sec:AnalysisStrategyGGHSF}

Similarly to the DF $\Pg\Pg\PH$-tagged analysis described in
Section~\ref{sec:AnalysisStrategyGGHOF}, an analysis targeting $\Pg\Pg\PH$ in
the SF $\EE$ and $\MM$ channels is performed.
The main challenge in this final state is
the large DY background contribution. In order to
control it, a BDT is trained to build a discriminator, called
DYMVA, to identify DY events.

A categorization based on the \pt
of the subleading lepton is introduced to better control the
nonprompt lepton background, and a categorization in the number of jets is used to
control the top quark backgrounds. The full list of event requirements is shown in
Table~\ref{Tab:AnalysisStrategy:selections:SF012}.

This is an event-counting analysis, and the event requirements are chosen to maximize the expected signal significance in each category.
The DY background estimations in these channels are based exclusively on control samples in data, as described in Section~\ref{backgrounds}.

\subsection{Associated \texorpdfstring{$\PW\PH$}{WH} production with three leptons in the final state}\label{sec:AnalysisStrategyWH3l}

The three-lepton $\PW\PH$-tagged analysis selects events that have the leading lepton with ${\pt{_{1}}>25\GeV}$, the subleading lepton with ${\pt{_{2}}>20\GeV}$,
and the trailing lepton with $\pt{_{3}}>15\GeV$. Events with a fourth lepton with $\pt>10\GeV$ are discarded. A veto is applied to events with SF lepton pairs of opposite charge that are compatible with coming from the decay of a $\PZ$ boson.
Events containing jets with $\pt>30\GeV$ or $\PQb$-tagged jets with
$\pt>20\GeV$ are also vetoed, to suppress the \ttbar background.
The azimuthal angle between \ptvecmiss and the three-lepton system \pt, $\Delta\phi(\ell\ell\ell,\ptvecmiss)$, is used to reduce the contamination of nonprompt lepton backgrounds.
The rest of the three-lepton $\PW\PH$-tagged selection is in common with the other categories.
These requirements are summarized in Table~\ref{Tab:AnalysisStrategy:selections:wh3l}.

The events are further divided into two categories: same-sign SF (SSSF) lepton pairs, ${\Pgm^{\pm}\Pgm^{\pm}{\Pe}^{\mp} / {\Pe}^{\pm}{\Pe}^{\pm}\Pgm^{\mp}}$,
and opposite-sign SF (OSSF) lepton pairs, $\Pgm^{\mp}\Pgm^{\pm}{\Pe}^{\mp}/{\Pe}^{\mp}{\Pe}^{\pm}\Pgm^{\mp}$.
The two selections have different signal-over-background ratios, with the SSSF being the purest of the two.
The main background contribution in both cases is the contamination from nonprompt leptons. In the OSSF category, events are required to have $\ptmiss > 50\GeV$ to reduce the DY background.

\begin{figure*}[!t]
\centering
\includegraphics[width=\cmsFigWidthAlt]{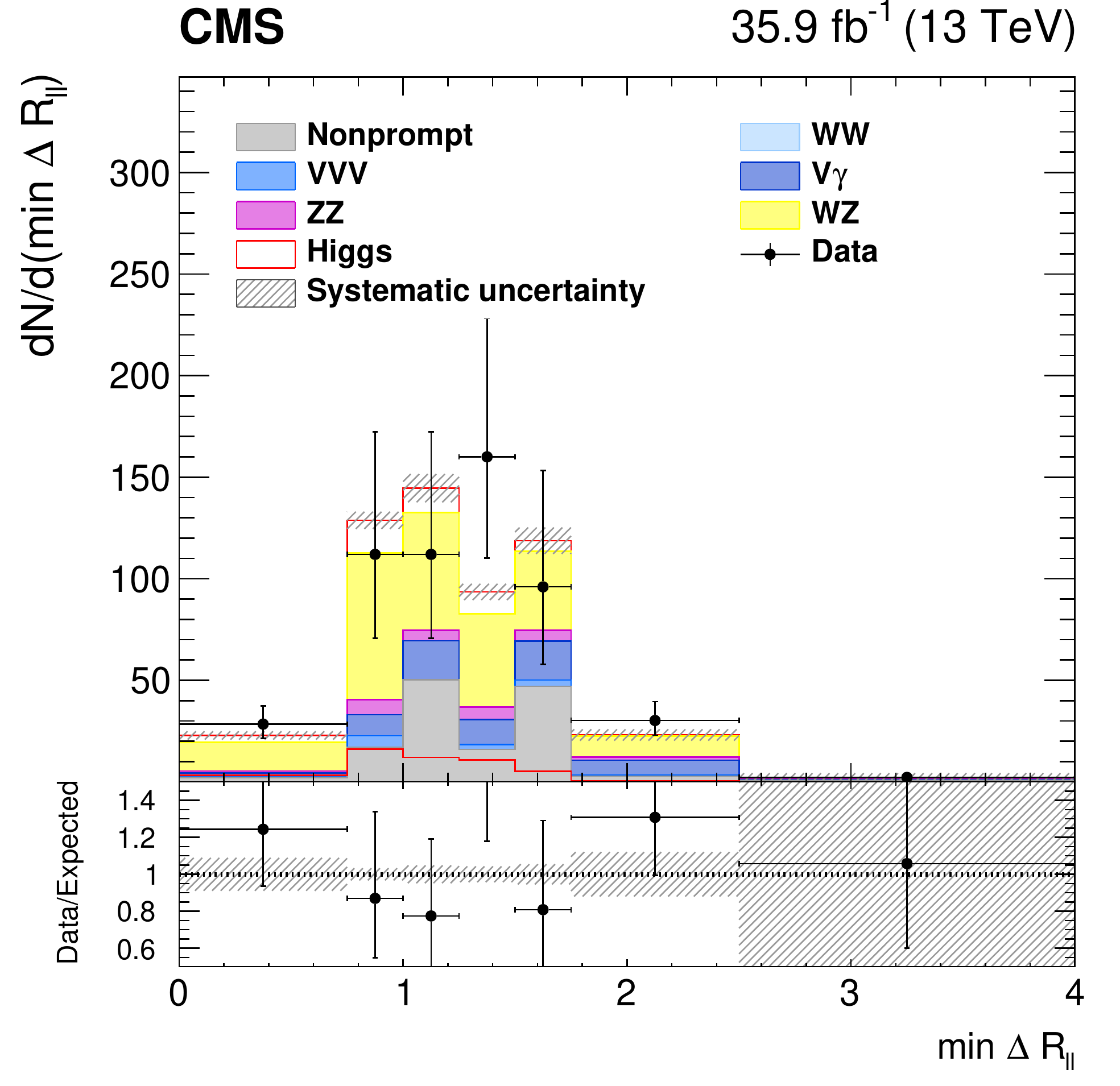}
\includegraphics[width=\cmsFigWidthAlt]{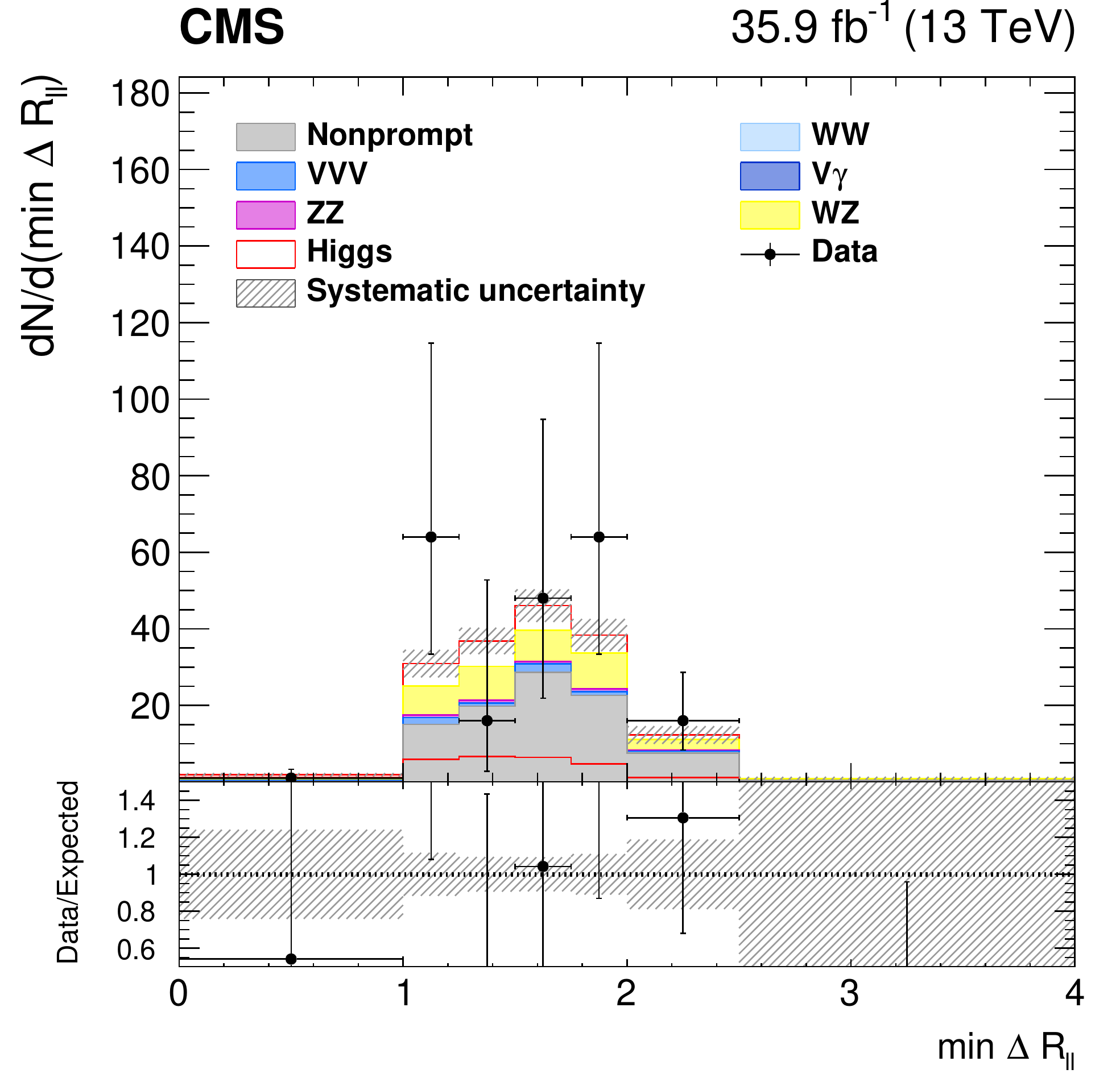}
\caption{ Postfit $\Delta R_{\ell\ell}$ distribution for events in the three-lepton
    $\PW\PH$-tagged category, split into the OSSF (left) and SSSF (right)
    subcategories.
    \label{fig:wh3l}
    }
\end{figure*}

The analysis is based on the minimum $\Delta R$ between oppositely charged leptons. The distribution of this variable is presented in Fig.~\ref{fig:wh3l}, separately for the SSSF and OSSF categories.

\subsection{Associated \texorpdfstring{$\PZ\PH$}{ZH} production with four leptons in the final state}\label{sec:AnalysisStrategyZH4l}

The $\PZ\PH$ final state is targeted by requiring exactly four isolated leptons with tight identification criteria and zero
total charge, and large \ptmiss from the undetected
neutrinos. The major background processes are $\PZ\PZ$ and $\ttbar\PZ$ production.

Among the four leptons, the pair of SF leptons with opposite charge, and with the invariant mass closest
to the $\PZ$ boson mass, is chosen as the $\PZ$ boson candidate. The remaining dilepton system, denoted as $\mathrm{X}$, can be either SF or DF.
Events are therefore divided into two categories, distinguishing
between the cases in which the $\mathrm{X}$ candidate contains two DF leptons (XDF) or two SF leptons (XSF), as shown in Table~\ref{Tab:AnalysisStrategy:selections:zh4l}.

The signal fraction is equally distributed in the two regions. In the XSF region,
$\PZ\PZ$, DY, and $\ttbar\PZ$ production are the major background sources, while in the XDF region,
$\ttbar\PZ$ and $\PZ\PZ$ backgrounds are dominant. Backgrounds with two $\PZ$ bosons fall
predominantly into the XSF region, and enter the XDF selection only through the leptonic
decays of the $\tau$ leptons. This makes the XDF region much cleaner than the XSF one.

Given the low expected signal yields in the XDF and XSF categories, the result
in this case is extracted from event-counting in each category.

\section{Background estimation\label{backgrounds}}

\subsection{Nonprompt lepton background}

Events in which a single $\PW$ boson is produced in association with jets may populate the signal region when a jet is misidentified as a lepton.
These events contain a genuine lepton and \ptmiss from the $\PW$ boson decay as well as a second nonprompt lepton from a misidentified jet, likely arising from a $\PB$ hadron decay. A similar background
arises from semileptonic decays of top quark pairs, especially in the 1- and 2- jets categories. At a lower rate,
multijet production and fully hadronic top quark pair decays also contribute. These backgrounds are
particularly important for events with low-\pt leptons and low \mll, and hence in the signal region of the analysis.

The nonprompt lepton background is suppressed by the identification and isolation requirements imposed on the electrons and muons, while the remaining
contribution is estimated directly from data.
A control sample is defined using events in which one lepton passes the standard lepton identification and isolation
criteria and another lepton candidate
fails these criteria but passes a looser selection, resulting in a sample of ``pass-fail'' lepton pairs.
The pass-fail sample is dominated by nonprompt leptons.
The efficiency ($\epsilon{_\mathrm{misID}}$) for a jet that satisfies this looser selection
to pass the standard selection is estimated directly
from data in an independent sample dominated by events with nonprompt leptons from multijet processes.
The contamination of prompt leptons from electroweak processes in such a sample is removed using the
simulation. The uncertainty from this subtraction is propagated to
$\epsilon{_\mathrm{misID}}$.
The efficiency $\epsilon{_\mathrm{misID}}$ is parameterized as a function of the \pt and
$\eta$ of the leptons, and is used to weight the events in the pass-fail sample by
$\epsilon{_\mathrm{misID}}/(1-\epsilon{_\mathrm{misID}})$, to obtain the estimated contribution from this background in
the signal region.
The contamination of prompt leptons in the ``pass-fail'' sample is corrected
for using their probability to pass the standard selection given that they pass the looser selection, as measured in a Drell--Yan data control sample.
The systematic uncertainty associated with the determination of $\epsilon{_\mathrm{misID}}$ is dominant and arises from the dependence of
$\epsilon{_\mathrm{misID}}$ on the composition of the jet that is
misidentified as a lepton.
Its impact is estimated in two independent ways, which are combined to yield a conservative result.
First, a closure test performed on simulated \wjets events with $\epsilon{_\mathrm{misID}}$ estimated from simulated QCD multijet events provides an overall normalization uncertainty.
Second, a shape uncertainty is derived by varying the jet \pt threshold in
the differential measurement of $\epsilon{_\mathrm{misID}}$ in bins of the
$\eta$ and \pt of the lepton.
The threshold is varied by a quantity that reflects the difference in the fake
lepton \pt spectrum between \wjets and \ttbar events.
The total uncertainty in $\epsilon{_\mathrm{misID}}$, including the statistical precision of the control sample, is about 40\%.
This uncertainty fully covers any data/simulation differences in control regions in which two same-sign leptons are requested.

\subsection{Top quark background}

Background contamination from single top quark processes, in particular t$\PW$ associated production, and from \ttbar production,
arises because of the inefficiency of $\PQb$ jet identification and the relatively large top quark cross sections at
13\TeV. The shapes of the top quark background distributions in the various categories are obtained from simulation, taking into account the measured $\PQb$ jet identification inefficiencies. The normalizations are obtained from control regions enriched in top quark events.
The background estimation is obtained separately
for the 0-, 1- and 2-jet $\Pg\Pg\PH$-tagged categories, the 2-jet VBF- and V$\PH$-tagged categories,
and for DF and SF final states.

The control region for the 0-jet $\Pg\Pg\PH$-tagged category is defined the same way as the signal region, except for the requirement that at least one jet
with $20<\pt<30\GeV$ is identified as a $\PQb$ jet by means of the $\PQb$ tagging algorithm.
For the 1-jet $\Pg\Pg\PH$-tagged top quark enriched region, exactly one jet with $\pt>30\GeV$ identified as a $\PQb$ jet is required.
In the 2-jet top quark enriched regions (either $\Pg\Pg\PH$-, V$\PH$-, or VBF-tagged), two jets with $\pt>30\GeV$
must be present in the event and at least one has to be identified as a $\PQb$ jet.
To reduce other backgrounds in the top quark control regions, the dilepton mass is
required to be higher than 50\GeV.
The derived scale factors are shown in Table~\ref{tab:sftop}.
The normalization of the top quark background in the three- and four-lepton categories is taken from simulation with its NNLO cross section uncertainty.
\begin{table}[htbp]
\small
\topcaption{
    Data-to-simulation scale factors for the top quark background normalization in seven different control regions.
}\label{tab:sftop}
\centering
\begin{tabular}{ccc}
\hline
Final state & Category  & Scale factor   \\
\hline
\multirow{5}{*}{DF} 		  & 0-jet $\Pg\Pg\PH$-tagged             &  $0.94 \pm 0.05$ \\
                                  & 1-jet $\Pg\Pg\PH$-tagged             &  $0.94 \pm 0.03$ \\
                                  & 2-jet $\Pg\Pg\PH$-tagged             &  $0.98 \pm 0.02$ \\
                                  & 2-jet V$\PH$-tagged{\phantom{x}} &  $0.98 \pm 0.03$ \\
                                  & 2-jet VBF-tagged             &  $1.01 \pm 0.04$ \\
[\cmsTabSkip]
\multirow{2}{*}{SF}     	  & 0-jet $\Pg\Pg\PH$-tagged             &  $1.03 \pm 0.06$ \\
                                  & 1-jet $\Pg\Pg\PH$-tagged             &  $0.98 \pm 0.02$ \\
\hline
\end{tabular}
\end{table}

The top quark \pt in \ttbar events is reweighted in simulated samples in order to have a better description of the \pt distribution observed in data, as described in previous CMS analyses~\cite{Khachatryan:2016mnb}. The difference between applying this reweighting, or not, is taken as a systematic shape uncertainty.
The theoretical uncertainty related to the single top quark and \ttbar cross
sections is also taken into account. It is evaluated by varying the ratio
between the single top quark and \ttbar cross section by its uncertainty, which is 8\% at 13\TeV~\cite{Chatrchyan:2013iaa}.
A 1\% theoretical uncertainty arising from PDF uncertainties and QCD scale
variations affects the uncertainty on the signal region to control region ratio.
All the experimental uncertainties described in Section~\ref{systematics} are also included as uncertainties on the top quark background shape.

\subsection{Drell--Yan background}\label{sec:DYbackground}

The $\mathrm{DY}\to\TT$ background is relevant for DF
categories and, like the signal, populates the low-\mt and low-\mll phase space.
The kinematic variables of this background are predicted by the simulation after reweighting the $\PZ$ boson \pt spectrum to match the distribution measured in the data.
The normalization is estimated in data control regions by selecting events
with $\mt<60\GeV$ and $30 < \mll < 80\GeV$.
Normalization scale factors are extracted, separately for the
0-, 1-, 2-jet $\Pg\Pg\PH$-tagged, the 2-jet VBF- and V$\PH$-tagged categories, and are shown in Table~\ref{tab:sfdy}.

\begin{table}[htbp]
\small
\topcaption{
    Data-to-simulation scale factors for the $\mathrm{DY}\to\TT$ background
    normalization in the DF control regions.
}\label{tab:sfdy}
\centering
\begin{tabular}{ccc}
\hline
Final state & Category  & Scale factor   \\
\hline
\multirow{5}{*}{DF} 		  & 0-jet $\Pg\Pg\PH$-tagged             &  $0.94 \pm 0.06$ \\
                                  & 1-jet $\Pg\Pg\PH$-tagged             &  $1.02 \pm 0.05$ \\
                                  & 2-jet $\Pg\Pg\PH$-tagged             &  $0.99 \pm 0.09$ \\
                                  & 2-jet V$\PH$-tagged{\phantom{x}} &  $0.99 \pm 0.13$ \\
                                  & 2-jet VBF-tagged             &  $1.04 \pm 0.16$ \\
\hline
\end{tabular}
\end{table}
The effect of missing higher-order corrections in the DY simulation is estimated by varying the renormalization and factorization scales by a factor of two up and down. This effect is treated as a shape uncertainty and amounts to 1--2\% in the DY yield.
A 2\% theoretical uncertainty arising from PDF
uncertainties and scale variations affects the uncertainty on the signal
region to control region ratio.
All experimental uncertainties described in Section~\ref{systematics} are considered as shape uncertainties for this background process.

In the SF categories, a dominant source of background is
$\mathrm{DY}\to\EE$ and $\mathrm{DY}\to\MM$.
The contribution of the DY background outside the $\PZ$ boson mass region (dubbed the \textit{out} region, which corresponds to the signal
region of the analysis) is estimated by
counting the number of events in the $\PZ$ boson mass region in data (\textit{in} region),
subtracting the non-Z-boson contribution from it, and scaling the yield by a ratio
$R_\mathrm{out/in}$. This ratio is defined as the fraction of events outside and inside the $\PZ$ boson mass
region in Monte Carlo (MC) simulation, $R_\mathrm{out/in} = N_\mathrm{out}^\mathrm{MC}/N_\mathrm{in}^\mathrm{MC}$.

 The $\PZ$ boson mass region is defined as $\abs{\mll-m_{\PZ}}<7.5\GeV$.
 Such a tight mass window is chosen to reduce the non-Z-boson background
contributions, which can be split into two categories.
The first one is composed of the background processes, such as top quark pair and $\PWp\PWm$ production, with equal decay rates into the four lepton-flavor final states ($\Pe\Pe$, $\Pe\Pgm$, $\Pgm\Pe$, and $\Pgm\Pgm$).
Their contributions to the $\PZ$ boson mass region in data, $N_{\ell\ell}^\mathrm{background|in}$, can be estimated from the
number of events in the $\Pe^{\pm}\Pgm^{\mp}$ final state, $N_{{\Pe}\mu}^\mathrm{in}$, applying a correction factor that accounts for the differences in the detection efficiency between electrons and muons ($k_{\Pe\Pe}$ and $k_{\Pgm\Pgm}$):
\begin{equation}
N_{\ell\ell}^\mathrm{background|in}=\frac{1}{2}k_{\ell\ell}(N_{{\Pe}\Pgm}^\mathrm{in}-N_{{\Pe}\Pgm}^\mathrm{in}(\mathrm{VV})),
\label{Rinout}
\end{equation}
where $\ell\ell$ stands for $\Pe\Pe$ or $\Pgm\Pgm$.
$N_{{\Pe}\Pgm}^\mathrm{in}(\mathrm{VV})$ is the number of events, estimated
from simulation, arising from $\PW\PZ$ and $\PZ\PZ$ decays and contributing to the
$\Pe\Pgm$ final state.
The factor of $1/2$ comes from the relative branching fraction between the $\ell\ell$ and $\Pe\Pgm$ final states.
The second category is composed of background processes, such as $\PW\PZ$ and
$\PZ\PZ$ (denoted as VV) production, with subsequent decay mostly into SF
final states via the on-shell $\PZ$ boson, which are determined from
simulation. The number of events arising from these background processes
contributing to the same flavor final state is denoted as $N_\mathrm{\ell\ell}^\mathrm{in}(\mathrm{VV})$.

Finally, the number of DY events in the signal region is estimated from the number of events in the SF final state, $N_\mathrm{\ell\ell}^\mathrm{in}$, separately for electrons and muons according to the following formula:
\begin{equation}
\label{NRinout}
N_\mathrm{\PZ\to\ell\ell}^\mathrm{out} = R_\mathrm{out/in} \left(
N_\mathrm{\ell\ell}^\mathrm{in}-N_{\ell\ell}^\mathrm{background|in}-N_\mathrm{\ell\ell}^\mathrm{in}(\mathrm{VV})
\right).
\end{equation}
The difference of the $R_\mathrm{out/in}$ values from the data and simulation is taken as a systematic uncertainty, and amounts to 10--25\%.

\subsection{The \texorpdfstring{$\PW\PZ$}{WZ} and \texorpdfstring{$\mathrm{W}\PGg^{*}$}{Wg*} background}

The $\PW\Pgg^{*}$ EW production is included in the simulation as part of the $\PW\PZ$ production, and the two processes are separated using a 4\GeV threshold on the $\PZ/\Pgg^*$ mass at the generator level.
For the final states with two leptons, the $\PW\PZ$ and $\PW\Pgg^{*}$ processes may contribute to the signal region whenever
one of the three leptons is not identified. Therefore, it is important to observe the process in data to validate the simulation.

The yield of the $\PW\PZ$ background is measured in data by selecting events with
three isolated leptons, two electrons and one muon ($\Pe\Pe\Pgm$), or two muons and one electron ($\Pgm\Pgm\Pe$). The SF lepton pair is identified as the $\PZ$ boson candidate, and its invariant mass is required to be within the $\PZ$ boson mass window defined in Section~\ref{sec:DYbackground}. This phase space is used to derive a scale factor for the $\PW\PZ$ simulation, which
is found to be $1.14 \pm 0.18$, from the weighted average of the scale factors
in the $\Pe\Pe\Pgm$ and $\Pgm\Pgm\Pe$ regions with their statistical
uncertainties.

A $\PW\Pgg^{*}$-enriched control region is defined by selecting events with
two muons with invariant mass below 4\GeV, likely arising from a $\Pgg^{*}$
decay, and a third isolated electron or muon passing a tight identification
requirement. The dimuon invariant mass region close to the J$/\psi$ resonance
mass is discarded. This control region is used to derive a scale factor for
the $\PW\Pgg^{*}$ simulation, which is found to be $0.9 \pm 0.2$, with the
uncertainty coming from the event counts in the $\Pgm\Pgm\Pe$ and $\Pgm\Pgm\Pgm$ samples.

All experimental uncertainties described in Section~\ref{systematics} are
considered as shape and yield uncertainties for the $\PW\PZ$ and $\PW\Pgg^{*}$
background determination. Moreover the effects of scale and PDF uncertainties on the normalization (3\% from scale variations and 4\% for PDF variations) and acceptance (3\%) are included.

\subsection{Nonresonant \texorpdfstring{$\PW\PW$}{WW} and other backgrounds}

The nonresonant $\PW\PW$ background populates the entire two-dimensional phase space in \mll and \mt, while
the Higgs boson signal is concentrated at low \mll values, and \mt values around the Higgs boson mass.
The yield of this background is hence estimated directly from the fit procedure, separately for each category.
The derived scale factors are shown in Table~\ref{tab:sfww}.

\begin{table}[htbp]
\small
\topcaption{
    Scale factors for the nonresonant $\PW\PW$ background normalization.
}\label{tab:sfww}
\centering
\begin{tabular}{ccc}
\hline
Final state & Category  & Scale factor   \\
\hline
\multirow{5}{*}{DF} & 0-jet $\Pg\Pg\PH$-tagged              & $1.16 \pm 0.05$ \\
                                  & 1-jet $\Pg\Pg\PH$-tagged              & $1.05 \pm 0.13$ \\
                                  & 2-jet $\Pg\Pg\PH$-tagged              & $0.8 \pm 0.4$ \\
                                  & 2-jet V$\PH$-tagged{\phantom{x}}  & $0.6  \pm 0.6$  \\
                                  & 2-jet VBF-tagged              & $0.5  \pm 0.5$ \\
[\cmsTabSkip]
\multirow{2}{*}{SF}      & 0-jet $\Pg\Pg\PH$-tagged              & $1.13 \pm 0.07$ \\
                                  & 1-jet $\Pg\Pg\PH$-tagged              & $1.03 \pm 0.18$ \\
\hline
\end{tabular}
\end{table}

In the $\qqbar\to\PW\PW$ process, the $\pt^{\PW\PW}$ spectrum in simulation is reweighted to match the resummed calculation~\cite{Meade:2014fca,Jaiswal:2014yba}.
The modeling of the shape uncertainties related to missing higher orders is done in two pieces: the first varies the factorization and renormalization scales by a factor of two up and down and takes the envelope; the second independently varies the resummation scale by a factor of two up and down.
The cross section of the gluon-induced $\PW\PW$ process is scaled to NLO accuracy and the uncertainty on this $K$ factor is 15\%~\cite{Passarino:2013bha}.
In categories with at least two jets, the EW $\PW\PW$ production is also taken into account. The theoretical uncertainty in the LO cross section of this process amounts to 11\%, and is estimated by varying the renormalization and factorization scales by a factor of two up and down, including also the effect of PDF variations.

The $\PW\PZ$ and $\PZ\Pgg^{*}$ backgrounds in the three-lepton $\PW\PH$-tagged
analysis are estimated using dedicated control regions from which the scale factors of
$1.09\pm{}0.06$ and $1.61 \pm 0.18$, respectively, are derived. The $\PZ\PZ$ background in the four-lepton $\PZ\PH$-tagged analysis is also estimated using a control region from which a scale factor
of $0.96 \pm 0.07$ is derived.

All remaining backgrounds from diboson and triboson production are estimated
according to their expected theoretical cross sections and the shape is taken from simulation.

\section{Statistical procedure and systematic uncertainties\label{systematics}}

The statistical methodology used to interpret subsets of data selected for the $\PH\to\PW\PW$ analysis
and to combine the results from the independent categories
has been developed by the ATLAS and CMS Collaborations in the context of the LHC Higgs Combination Group.
A general description of the methodology can be found in Ref.~\cite{LHC-HCG}.

The number of events in each category and in each bin of the discriminant distributions
used to extract the signal is modeled as a Poisson random variable,
with a mean value that is the sum of the contributions from the processes under consideration.
Systematic uncertainties are represented by individual nuisance parameters
with log-normal distributions. The uncertainties affect the overall
normalizations of the signal and backgrounds,
as well as the shapes of the predictions across the distributions of the observables.
Correlations between systematic uncertainties in different categories
are taken into account.

The various control regions described in Section~\ref{backgrounds} are used
to constrain individual backgrounds and are included in the fit in the form of
single bins, representing the number of events in each of the control regions.

The remaining sources of systematic uncertainties of experimental and theoretical nature are described below.
Effects due to the experimental uncertainties are estimated by scaling or smearing the targeted variable in the simulation
and recalculating the analysis results.
All experimental sources of systematic uncertainty, except for the integrated
luminosity, have both a normalization and a shape component. The following
experimental uncertainties are taken into account:
\begin{itemize}
\setlength\itemsep{0em}
\item The uncertainty in the measured luminosity, which is 2.5\%~\cite{CMS-PAS-LUM-17-001}.
\item The trigger efficiency uncertainty associated with the combination of single-lepton and dilepton triggers, which is 2\%~\cite{Khachatryan:2016bia}.
\item The uncertainties in the lepton reconstruction and identification efficiencies, which vary within 2--5\% for electrons~\cite{Khachatryan:2015hwa}
      and 1--2\% for muons~\cite{Chatrchyan:2012xi}, depending on \pt and $\eta$.
\item The muon momentum and electron energy scale and resolution uncertainties, which amount to 0.6--1.0\% for electrons
      and 0.2\% for muons.
\item The jet energy scale uncertainties, which vary in the range 1--13\%, depending on the \pt and $\eta$ of the jet~\cite{Khachatryan:2016kdb}.
\item The \ptmiss resolution uncertainty includes the propagation of lepton and jet energy scale and resolution uncertainties to
\ptmiss, as well as the uncertainties on the energy scales of particles that are not clustered into jets, and the uncertainty on the amount of energy coming from pileup interactions.
\item The scale factors correcting the $\PQb$ tagging efficiency and mistagging rates, which are varied within their
      uncertainties. The associated systematic uncertainty, which varies between 0.5--1.0\%~\cite{Sirunyan:2017ezt}, affects, in an anticorrelated way, the top quark control regions and
      the signal ones.
\end{itemize}

The uncertainties in the signal and background production rates due
to the limited knowledge of the processes under study include several components, which are assumed to be
independent: the choices of PDFs and the strong coupling constant \alpS, the UE and parton shower model,
and the effects of missing higher-order corrections via variations of the renormalization
and factorization scales. As most of the backgrounds are estimated from control regions in data,
these theoretical uncertainties mostly affect the Higgs boson signal and they are implemented as normalization-only uncertainties unless stated otherwise.

The PDFs and \alpS uncertainties are further split between the
cross section normalization uncertainties computed by the LHC Higgs Cross Section Working Group~\cite{Heinemeyer:2013tqa}
for the Higgs boson signal and their effect on the acceptance~\cite{Butterworth:2015oua}.
The signal cross section normalization uncertainties amount to 3\% for the
$\Pg\Pg\PH$ and 2\% for the VBF Higgs boson production mechanism, between 1.6\% and 1.9\% for V$\PH$ processes, and 3.6\% for $\ttbar\PH$ production.
The acceptance uncertainties are less than 1\% for all production mechanisms.

The effect of missing higher order QCD corrections on the $\Pg\Pg\PH$
production mechanism is split into nine individual components as identified in Ref.~\cite{deFlorian:2016spz}, chapter I.4. Each component is propagated such that both the integrated effect and the correlations across different categories are properly taken into account. The overall effect on the $\Pg\Pg\PH$ cross section is about 10\%.
The effect of missing higher-order corrections in the VBF and V$\PH$ simulations is less than 1\%, while it amounts to about 8\% for the $\ttbar\PH$ simulation.

The UE uncertainty is estimated by varying the  CUET8PM1 tune in a range
corresponding to the envelope of the single tuned parameters post-fit
uncertainty,  as described in Section~\ref{section:mcdata}.
The dependence on the parton shower (PS) model is estimated by comparing samples
processed with different programs, as described in Section~\ref{section:mcdata}.
The effect on the expected $\Pg\Pg\PH$ signal yields after preselection is about 5\% for the UE tuning and
about 7\% for the PS description, and is partially accounted for by the lepton identification scale factors and uncertainties.
The remaining contribution is migration between jet categories and is
anticorrelated between the 0-jet category and the categories with jets.
Such effects are of the order of 15-25\% for the parton shower (VBF categories being the most
affected) and 5-17\% for UE (2-jet V$\PH$-tagged category being the most affected).
The anticorrelation between jet categories reduces
the impact of these uncertainties on the final results.

Finally, the uncertainties arising from the limited number of events in the simulated samples are included independently for each bin
of the discriminant distributions in each category.

\section{Results}\label{sec:Results}

The signal strength modifier ($\mu$), defined as the ratio between the measured signal
cross section and the SM expectation in the $\PH \to \PW\PW \to 2\ell 2\nu$ decay
channel, is measured by performing a binned maximum likelihood fit using
simulated binned templates for signal and background processes.

\begin{figure}[htbp]
\centering
\includegraphics[width=\cmsFigWidthSingle]{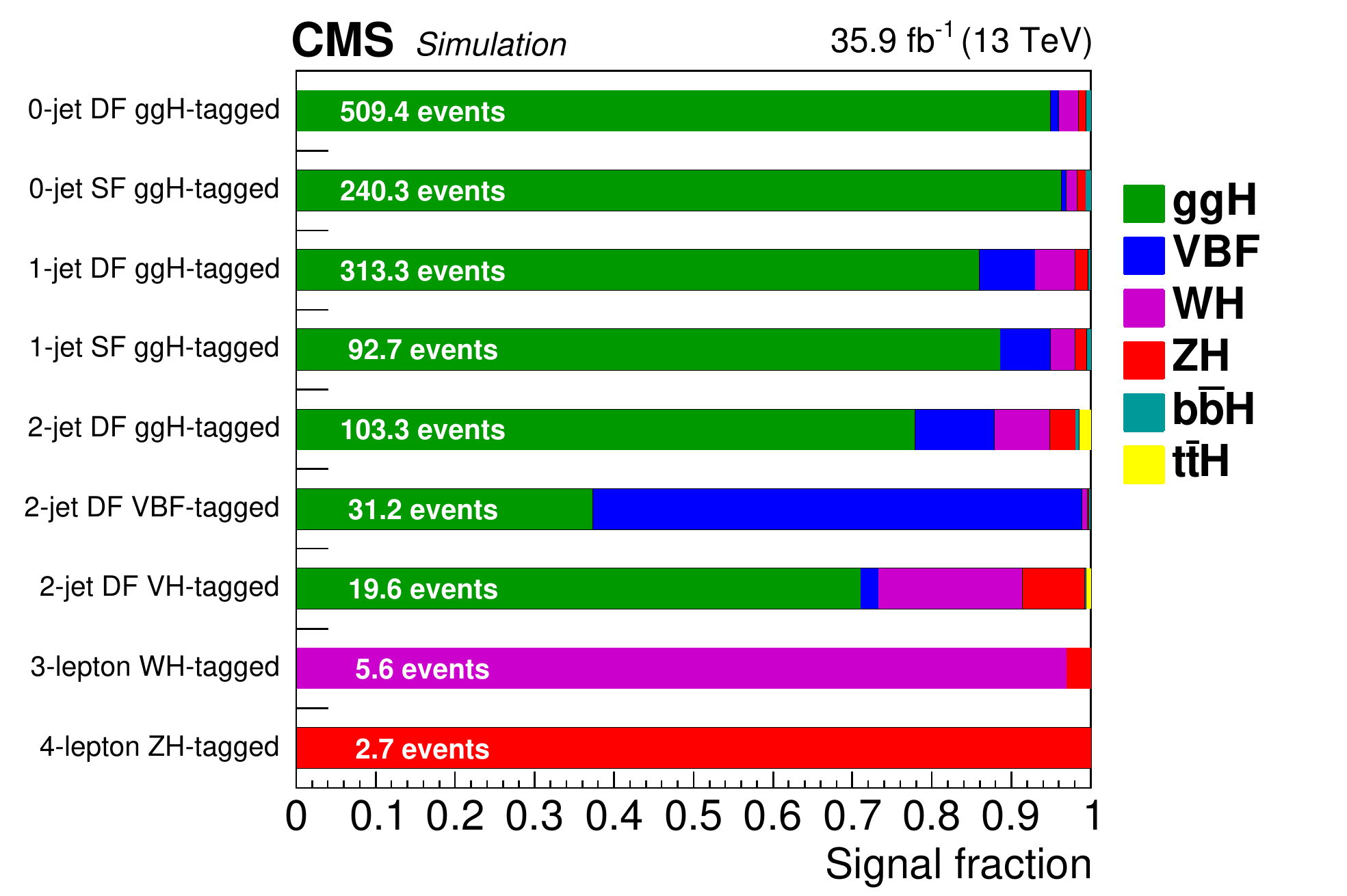}
\caption{
  Expected relative fraction of different Higgs boson production mechanisms in each category included in the combination, together with the expected signal yield.
}\label{fig:signal_fraction}
\end{figure}

The combined results obtained using all the individual analysis categories are
described in this section. A summary of the expected fraction of different
signal production modes in each category is shown in
Fig.~\ref{fig:signal_fraction}, together with the total number of expected
$\PH \to \PW\PW$ events. The chosen categorization proves effective in
tackling the different production mechanisms, especially $\Pg\Pg\PH$, VBF, and V$\PH$.
The measurements assume a Higgs boson mass of $m_{\PH}=125.09\GeV$, as reported in the ATLAS and CMS combined Higgs boson mass measurement~\cite{Aad:2015zhl}.
The results reported below show a very weak dependence on the Higgs boson mass
hypothesis, with the expected signal yield varying within 1\% when the signal
mass hypothesis is varied within its measured uncertainty.

The number of expected signal and background events, and the number of observed events in data, in each category after the full event selection are shown in Tables~\ref{tab:yields_01j} and \ref{tab:yields_2j34l}.
\begin{table*}[!t]
\small
\topcaption{
    Number of expected signal and background events and number of observed events in the 0- and 1-jet categories after the full event selection. Postfit event yields are also shown in parentheses, corresponding to the result of a simultaneous fit to all categories assuming that the relative proportions for the different production mechanisms are those predicted by the SM.
    The individual signal yields are given for different production mechanisms. The total uncertainty accounts for all sources of uncertainty in signal and background yields after the fit.}\label{tab:yields_01j}
\centering

\cmsTable{
\begin{tabular}{lr@{\hskip 0.1cm}r@{\hskip 0.8cm}r@{\hskip 0.1cm}r@{\hskip 0.8cm}r@{\hskip 0.1cm}r@{\hskip 0.8cm}r@{\hskip 0.1cm}r}
\hline

           & \multicolumn{2}{c}{\begin{tabular}{c} 0-jet DF \\ $\Pg\Pg\PH$-tagged \end{tabular}} &  \multicolumn{2}{c}{\begin{tabular}{c} 1-jet DF \\ $\Pg\Pg\PH$-tagged \end{tabular}} & \multicolumn{2}{c}{\begin{tabular}{c} 0-jet SF \\ $\Pg\Pg\PH$-tagged \end{tabular}} &  \multicolumn{2}{c}{\begin{tabular}{c} 1-jet SF \\ $\Pg\Pg\PH$-tagged \end{tabular}} \\
\hline
$\Pg\Pg\PH$		      	& 483.1 & (642.1)&       269.1 & (339.3)		&       231.2 & (324.6)       	&       82.0 & (92.8)       	\\
VBF     		& 5.6 & (7.4)         		&       22.1 & (29.4)  	        &       1.5 & (2.5)       	&       5.9 & (9.3)       	\\
$\PW\PH$   			&12.4 & (16.4)         		&       15.8 & (20.6)             &       3.3 & (4.3)      		&       2.9 & (3.8)       	\\
$\PZ\PH$         	   	&5.2 & (6.9)          		&       5.0 & (6.7)          	&       2.6 & (3.4)      		&       1.4 & (1.8)      		\\
$\ttbar\PH$         &$<$0.1 & ($<$0.1)        	&       0.2 & (0.2)               &       $<$0.1 & ($<$0.1)        	&         $<$0.1 & ($<$0.1)       \\
$\bbbar\PH$    	&   3.4 & (4.4)        		&       1.5 & (2.0)   	        &       1.7 & (2.3)      		&       0.5 & (0.7)      		\\
[\cmsTabSkip]
Signal 		        &  509 & (677)		&	313  & (398) 	 &	 240 & (337)	&	  93 & (108)	\\
$\pm$total unc.        &      & ($\pm$31) 	& 	     & ($\pm$19)& 	     &	($\pm$24) & 	     & ($\pm$13)  \\
[\cmsTabSkip]
$\PW\PW$	   		&7851 & (9088)      		&       3553 & (3727)         &       1596 & (1805)       	&       373 & (365)      	\\
Top quark	   		&2505 & (2422)      		&       5395 & (5224)         &       334 & (339)      		&       452 & (443)      	\\
Nonprompt  		& 1555 & (1006)      		&       781 & (482)           &       301 & (260)      	        &       111 & (97)      	\\
DY	   		&	154 & (154)   		&       283 & (302)           &       437 & (459)      		&       178 & (216)     	\\
$\mathrm{V\PZ}/\mathrm{V}\PGg^*$	&368 & (385)   		&       327 & (338)           &       101 & (104)      		&       43 & (43)      	\\
V$\PGg$  		&213 & (210)      		&       137 & (128)           &       23 & (26)      		&       17 & (19)      	\\
Other diboson  		&5.1 & (5.3)      		&       3.5 & (3.7)           &       9.3 & (9.4)      		&       2.0 & (2.1)      		\\
Triboson		&9.3 & (9.6)      		&       16 & (17)             &       1.2 & (1.2)      		&       1.3 & (1.3)      		\\
[\cmsTabSkip]
Background		& 12660  & (13280) 	&  10496  & (10222) 	&	  2803	& (3004) &	  1177  & (1186)	\\
$\pm$total unc.        &        & ($\pm$141)  & 	  & ($\pm$178) & 		& ($\pm$97)& 		& ($\pm$83)		\\
[\cmsTabSkip]
Data       		&	\multicolumn{2}{c}{13964}			&	\multicolumn{2}{c}{10591}			&	\multicolumn{2}{c}{3364}			&	\multicolumn{2}{c}{1308}			\\

\hline
\end{tabular}
}
\end{table*}
\begin{table*}[!t]
\small
\topcaption{
    Number of expected signal and background events and number of observed events in the 2-jet, 3-lepton, and 4-lepton categories after the full event selection. Postfit event yields are also shown in parentheses, corresponding to the result of a simultaneous fit to all categories assuming that the relative proportions for the different production mechanisms are those predicted by the SM.
    The individual signal yields are given for different production mechanisms. For the 3-lepton $\PW\PH$-tagged category, the ``Other diboson'' background includes mainly $\PW\PZ$ production, with a 10\% contribution from $\PZ\PZ$ events. For the 4-lepton $\PZ\PH$-tagged category, $\ttbar\PW$ and $\ttbar\PZ$ are included in the top quark process, while the ``Other diboson'' background mainly comes from $\PZ\PZ$ production. The total uncertainty accounts for all sources of uncertainty in signal and background yields after the fit.
    }\label{tab:yields_2j34l}
\centering

\cmsTable{
\begin{tabular}{lr@{ }r@{\hskip 0.8cm}r@{ }r@{\hskip 0.8cm}r@{ }r@{\hskip 0.8cm}r@{ }r@{\hskip 0.8cm}r@{ }r}
\hline

           & \multicolumn{2}{c}{\begin{tabular}{c} 2-jet DF \\ $\Pg\Pg\PH$-tagged \end{tabular}} &  \multicolumn{2}{c}{\begin{tabular}{c} 2-jet DF \\ VBF-tagged \end{tabular}} &  \multicolumn{2}{c}{\begin{tabular}{c} 2-jet DF \\ V$\PH$-tagged \end{tabular}} &  \multicolumn{2}{c}{\begin{tabular}{c} 3-lepton \\ $\PW\PH$-tagged \end{tabular}} &  \multicolumn{2}{c}{\begin{tabular}{c} 4-lepton \\ $\PZ\PH$-tagged \end{tabular}} \\
\hline
$\Pg\Pg\PH$                     &       80.4 &  (100.6)            &       11.6 & (14.6)             &       13.9 &  (17.4)             &         $<$0.1 & ( $<$0.1)                &          $<$0.1 &  ( $<$0.1)        \\
VBF                     &       10.3 &  (13.3)             &       19.2 & (24.5)             &       0.4  & (0.6)               &         $<$0.1 & ( $<$0.1)                &         $<$0.1 & ( $<$0.1)        \\
$\PW\PH$                      &       7.2  & (9.3)               &       0.2 & (0.2)               &       3.6  & (4.6)               &       5.4 & (7.2)                       &         $<$0.1 & ( $<$0.1)       \\
$\PZ\PH$                      &       3.3  & (4.3)               &        $<$0.1 & ( $<$0.1)         &       1.5  & (2.1)               &       0.2 & (0.2)                       &       2.7 & (3.5)   \\
$\ttbar\PH$         &       1.6  & (2.1)               &        $<$0.1 & ( $<$0.1)         &       0.1  & (0.2)               &         $<$0.1 & ( $<$0.1)                &         $<$0.1 & ( $<$0.1)   \\
$\bbbar\PH$         &       0.6  & (0.7)               &        $<$0.1 & (0.1)            &        $<$0.1 & ( $<$0.1)          &         $<$0.1 & ( $<$0.1)                &         $<$0.1 & ( $<$0.1)     \\
[\cmsTabSkip]
Signal                  &  103 & (130)           &       31 & (40)         &       20 & (25)        &         5.6 & (7.4)          &         2.7 & (3.5) \\
$\pm$total unc.        &  & ($\pm$16)          &       & ($\pm$3)       &      & ($\pm$3)       &        & ($\pm$0.7)         &        & ($\pm$0.3) \\
[\cmsTabSkip]
$\PW\PW$                      &       1048 &  (860)          &       69  & (46)             &       52  & (34)             &         $<$0.1 & ( $<$0.1)                &         $<$0.1 & ( $<$0.1)      \\
Top quark                    &       5197 &  (5187)         &       157 &  (158)           &       230 &  (229)           &         $<$0.1 & ( $<$0.1)                &           0.3 & (0.3)         \\
Nonprompt               &       359  & (305)           &       30  & (20)             &       42  & (37)             &        19   & (21)                  &         $<$0.1 & ( $<$0.1)       \\
DY                      &       110  & (112)           &       20  & (19)             &       29  & (30)             &         $<$0.1 & ( $<$0.1)                &         $<$0.1 & ( $<$0.1)     \\
$\mathrm{VZ}/\mathrm{V}\PGg^*$     &  136 & (137)           &       7.1  & (6.9)              &       11  & (10)             &         $<$0.1 & ( $<$0.1)                &         $<$0.1 & ( $<$0.1)       \\
V$\PGg$               &       59 & (53)             &       2.8 & (2.8)               &       4.2 & (4.6)               &       3.8 & (9.6)                       &         $<$0.1 &( $<$0.1)       \\
Other diboson           &        2.1 & (2.3)               &      0.3 & (0.3)               &       1.2 & (1.3)               &       32 & (37)                    &       13 &(13)     \\
Triboson                &       15 & (15)             &       0.3 & (0.3)               &       2.0 & (2.0)               &       2.1 & (2.1)                       &        0.4 &(0.4)    \\
[\cmsTabSkip]
Background              &  6926 & (6671)        &       287 & (253)      &       371 & (348)     &         57 & (70)               &         13.7 & (13.7) \\
$\pm$total unc.        &  & ($\pm$502)        &       & ($\pm$17)      &     &  ($\pm$37)     &         &($\pm$7)               &        & ($\pm$0.6) \\
[\cmsTabSkip]
Data                    &       \multicolumn{2}{c}{6802}                    &       \multicolumn{2}{c}{285}                     &       \multicolumn{2}{c}{386}                     &       \multicolumn{2}{c}{85}                              &       \multicolumn{2}{c}{15} \\
\hline
\end{tabular}
}

\end{table*}
Postfit event yields are also shown in parentheses, and correspond to the result of a simultaneous fit to all categories, assuming that the relative proportions of the different production mechanisms are those predicted by the SM.

\subsection{Signal strength modifiers}\label{sec:strength_and_signif}
\begin{figure}[!t]
\centering
\includegraphics[width=\cmsFigWidthSingle]{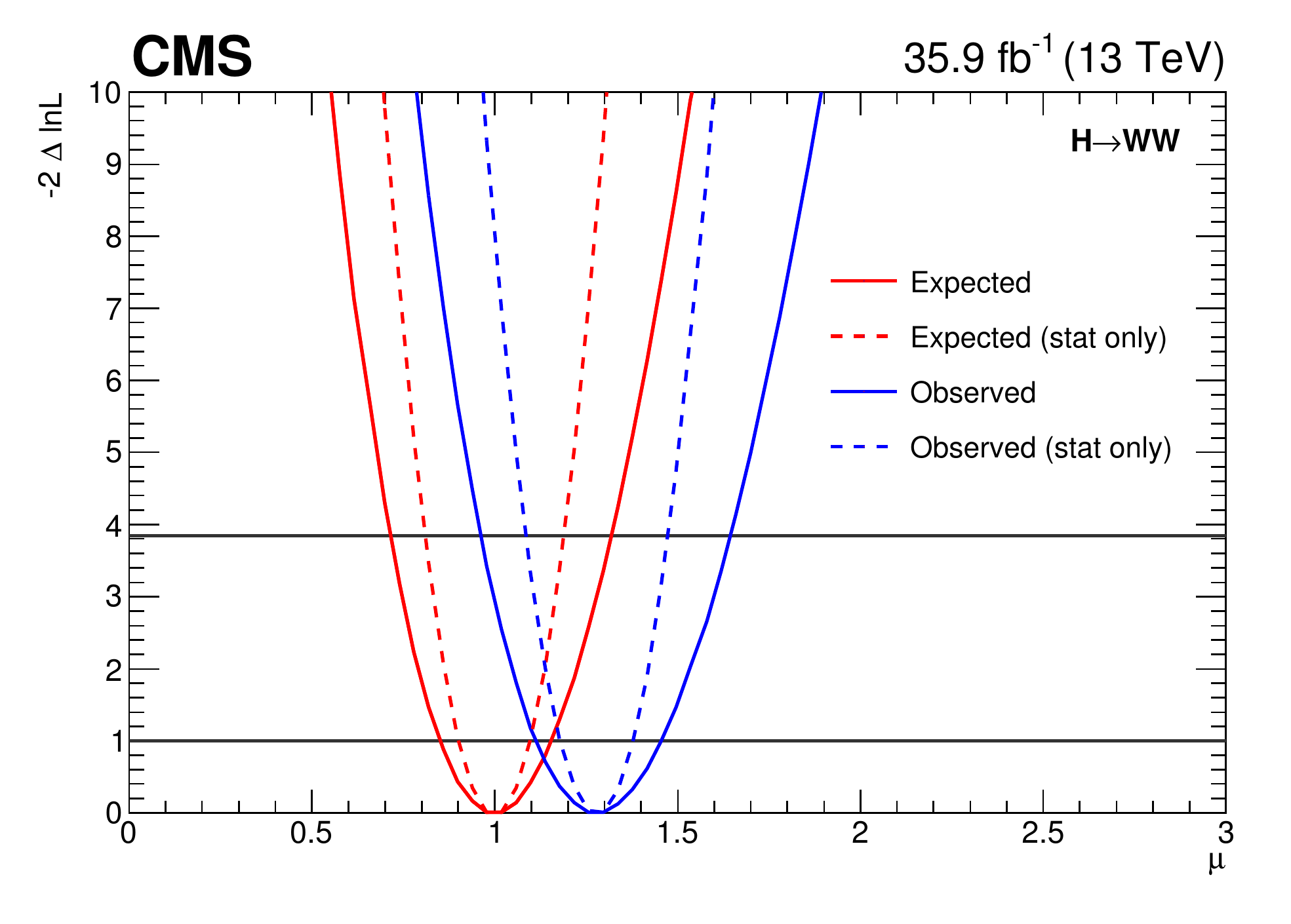}
\caption{
Observed and expected likelihood profiles for the global signal strength modifier. Dashed curves correspond to the likelihood profiles obtained including only the statistical uncertainty. The crossings with the horizontal line at $-2\Delta\ln L = 1~(3.84)$ define the 68 (95)\% \CL interval.
}
\label{fig:LHscan_comb}
\end{figure}

The signal strength modifier is extracted by performing a
simultaneous fit to all categories assuming that the relative proportions of
the different production mechanisms are the same as the SM ones. As such, the value of
$\mu$ provides an insight into the compatibility between this measurement
and the SM.
The combined observed signal strength modifier is:
\begin{equation}
\mu = 1.28 ^{+0.18}_{-0.17} = 1.28 \pm 0.10\stat\pm 0.11\syst^{+0.10}_{-0.07}\thy,
\end{equation}
\noindent where the statistical, systematic, and theoretical uncertainties are reported separately. The statistical component is estimated by fixing all the nuisance parameters to their best fit values and recomputing the likelihood profile. The breakdown of a given group of uncertainties (systematic or theoretical) is obtained by fixing all the nuisance parameters in the group to their best fit values, and recomputing the likelihood profile. The corresponding uncertainty is then taken as the difference in quadrature between the total uncertainty and the one obtained fixing the group of nuisance parameters.
The expected and observed likelihood profiles as functions of the signal
strength modifier are shown in Fig.~\ref{fig:LHscan_comb}, with the 68\% and
95\% confidence level (\CL) indicated.
The observed significance in the asymptotic approximation~\cite{Cowan:2010st} of the Higgs boson production for the combination of all categories is $9.1$ s.d., to be compared with the expected value of $7.1$ s.d. As such, this is the first observation of the Higgs boson decay to $\PW$
boson pairs with the CMS experiment.

\begin{table}[!b]
\small
\topcaption{
  Impact of the main systematic uncertainties on the signal strength $\mu$.
}\label{tab:syst_mu}
\centering
\begin{tabular}{ccc}
\hline
Type        & Source & Impact (\%) \\
\hline
\multirow{5}{*}{Theoretical} & Signal       & 7 \\
            & $\PW\PW$     & 2 \\
            & Top quark    & 2 \\
            & PS and UE    & 2 \\
            & Sample size of simulation data & 2 \\
[\cmsTabSkip]
\multirow{9}{*}{Experimental} & Electrons   & 5 \\
             & Luminosity  & 4 \\
             & Muons       & 3 \\
             & b-tagging   & 3 \\
             & Nonprompt   & 3 \\
             & Jets        & 2 \\
             & \ptmiss     & 2 \\
             & $\mathrm{VZ}/\mathrm{V}\PGg^*$ scale factor & 2 \\
             & DY SF scale factor & 2 \\
\hline
\end{tabular}
\end{table}

A breakdown of the impact on $\mu$ of the different systematic uncertainties
is shown in Table~\ref{tab:syst_mu}. The contributions of the normalizations
that are left floating in the fit enter the statistical error on $\mu$.

\begin{figure*}[!t]
\centering
\includegraphics[width=\cmsFigWidthAlt]{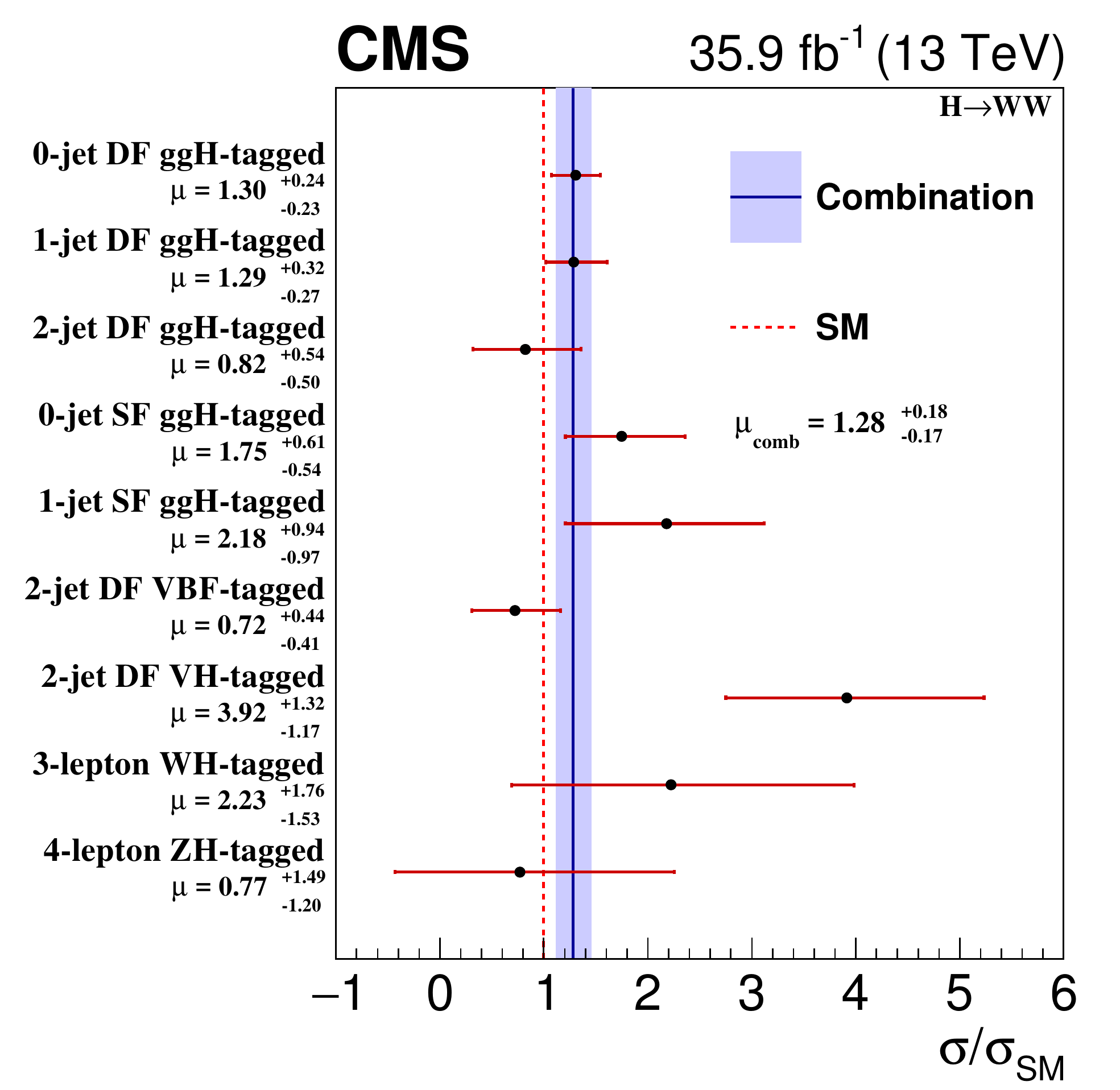}
\includegraphics[width=\cmsFigWidthAlt]{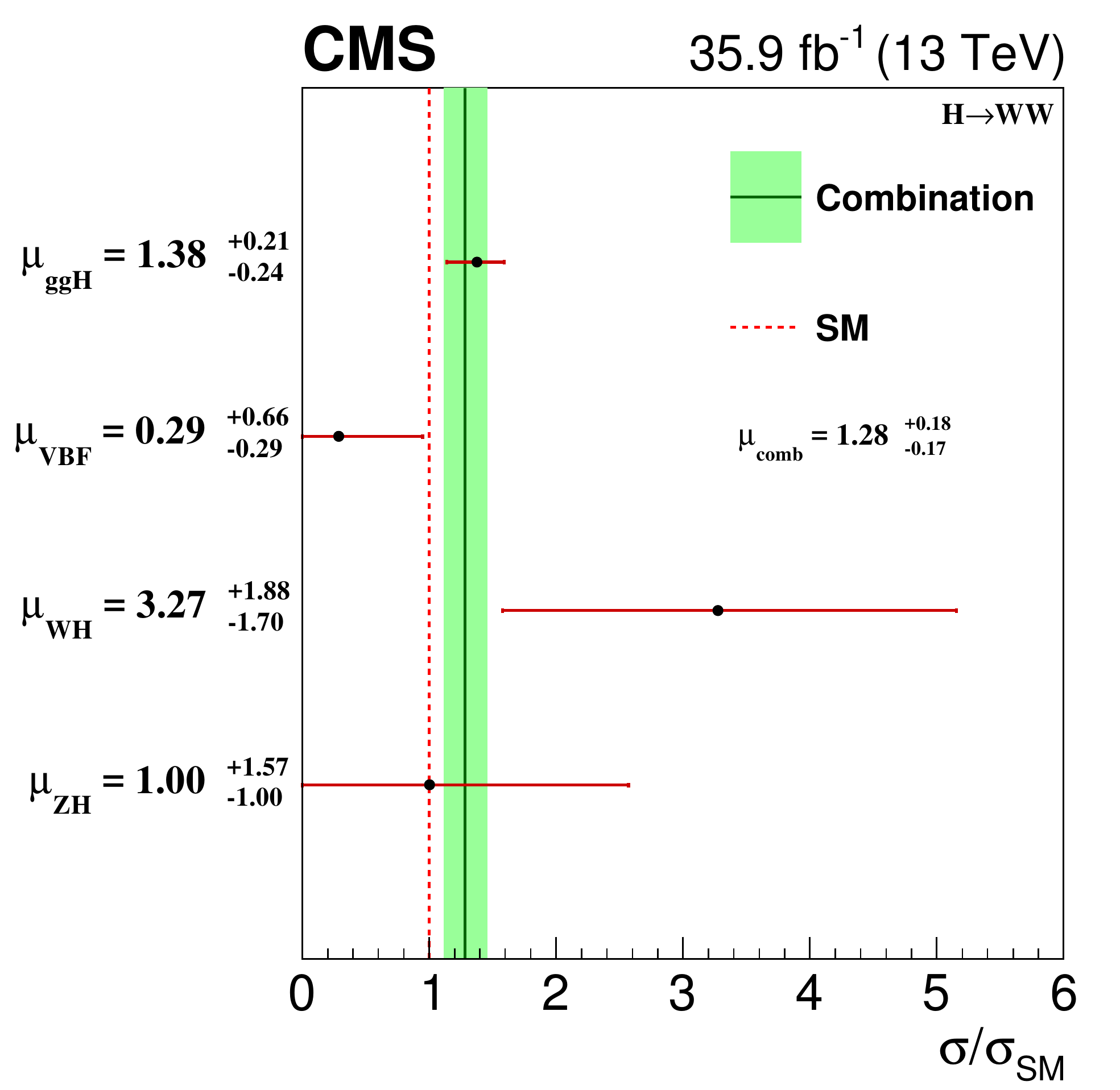}
\caption{
(Left) Observed signal strength modifiers for the categories used in the
combination. (Right) Observed signal strength modifiers corresponding to the main SM Higgs boson production mechanisms, for a Higgs boson with a mass of $125.09\GeV$. The vertical continuous line represents the combined signal strength best fit value, while the horizontal bars and the filled area show the 68\% confidence intervals. The vertical dashed line corresponds to the SM expectation.
}
\label{fig:mu_categories}
\end{figure*}

In order to assess the compatibility of the observed signal with the SM
predictions in each category of the analysis and to ascertain the
compatibility between the different categories, a simultaneous fit in which the
signal strength modifier is allowed to float independently in each category is performed.
The observed signal strength modifier for each category used in the
combination is reported in Fig.~\ref{fig:mu_categories} (left).
Results are generally consistent with unity, with the largest deviation
showing up in the 2-jet V$\PH$-tagged category (i.e., the category targeting the
associated production of a Higgs boson with a vector boson decaying hadronically).
The level of compatibility of the signal strength modifiers in each category with the combined signal strength modifier corresponds to an asymptotic $p$-value of 0.34.

Given the sensitivity of the analysis to various production mechanisms, a fit
is performed in which a different signal strength modifier is assigned to each
production mechanism, i.e., $\mu{_{\Pg\Pg\PH}}$, $\mu{_\mathrm{VBF}}$, $\mu{_{\PW\PH}}$,
and $\mu{_\mathrm{\PZ\PH}}$. A simultaneous fit to all categories is performed, and results are shown in Fig.~\ref{fig:mu_categories} (right).
The biggest deviation from unity is observed for the $\PW\PH$
production mechanism, which is probed mainly by the 2-jet V$\PH$-tagged and 3-lepton $\PW\PH$-tagged categories.
The level of compatibility of the signal strength modifiers associated with different production mechanisms with the combined signal strength modifier corresponds to an asymptotic $p$-value of 0.70.

\begin{figure}[!b]
\centering
\includegraphics[width=\cmsFigWidthSingle]{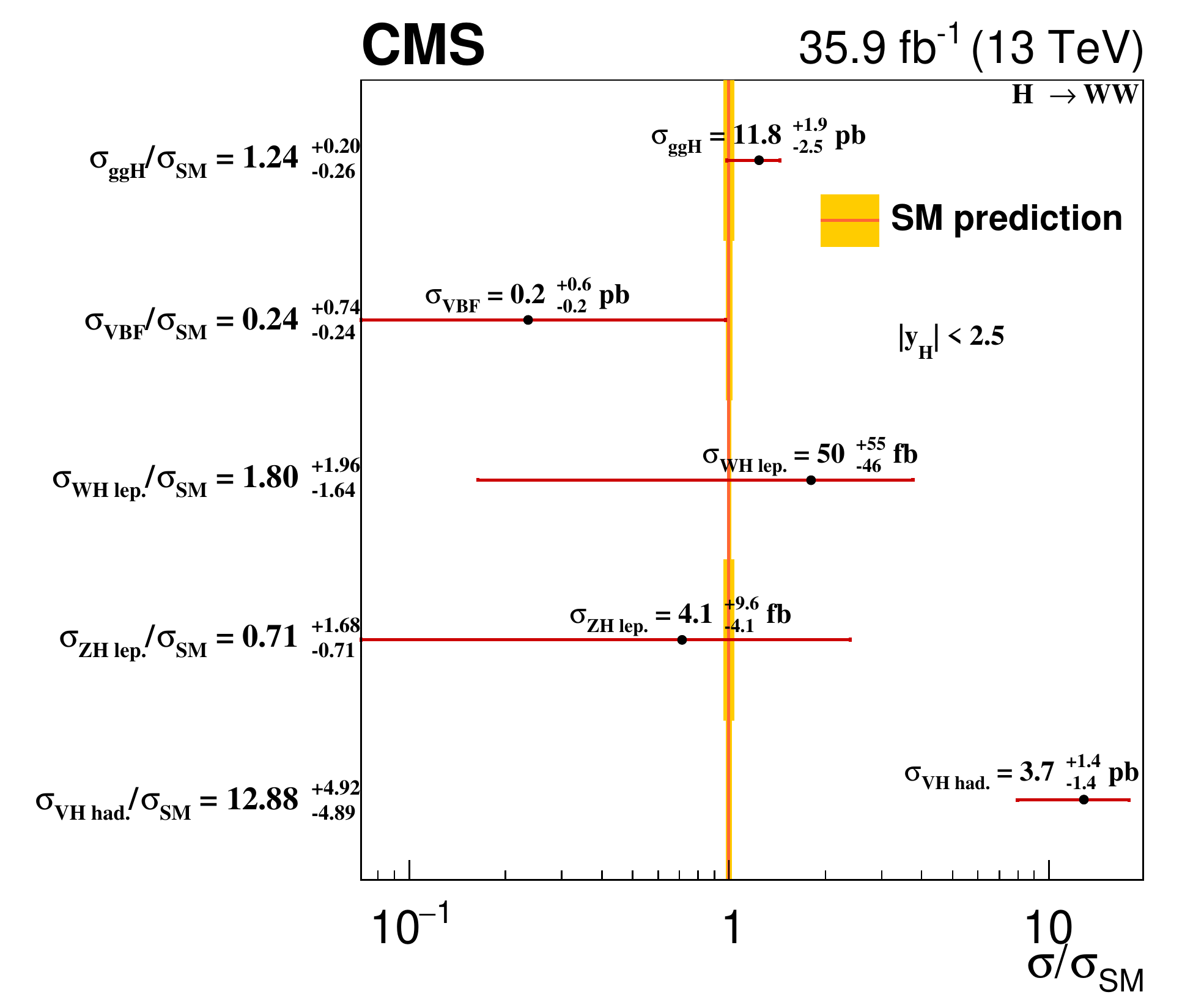}
\caption{
Observed cross sections and their ratio with the SM predictions for the main Higgs boson production modes. Cross section ratios are measured in a simplified fiducial
phase space defined by requiring $y_{\PH} < 2.5$, as specified in the ``stage-0'' simplified template cross section framework~\cite{deFlorian:2016spz}. The vertical line and band correspond to the SM prediction and associated theoretical uncertainty.
}
\label{fig:STXS}
\end{figure}

A similar simultaneous fit has been performed to measure the cross section
corresponding to five Higgs boson production mechanisms, using a
simplified fiducial phase space,
as specified in the ``stage-0'' simplified template cross section framework~\cite{deFlorian:2016spz}.
The cross sections corresponding to five Higgs boson production processes ($\sigma{_{\Pg\Pg\PH}}$, $\sigma{_\mathrm{VBF}}$, $\sigma{_\mathrm{\PW\PH~lep.}}$, $\sigma{_\mathrm{\PZ\PH~lep.}}$, $\sigma{_\mathrm{V\PH~had.}}$)
are measured requiring the generator-level Higgs boson rapidity to be $\abs{y_{\PH}} < 2.5$.
This analysis has a negligible acceptance for Higgs boson production above $\abs{y_{\PH}} = 2.5$.
The $\PH\to\PGt\PGt$ events are considered as background in this fit.
The measured cross sections and their ratio with the SM predictions, for the production
channels in which the analysis has sensitivity, are shown in Fig.~\ref{fig:STXS}.
The observed deviation of the $\sigma{_\mathrm{V\PH~had.}}$ process with respect to the SM prediction corresponds to an asymptotic $p$-value of 0.02,
and is driven by the excess of events already observed for $\mu{_\mathrm{\PW\PH}}$. Compared to the $\mu{_\mathrm{\PW\PH}}$ fit,
in this case the signal strength modifier for the hadronic decay of the associated $\PW$ boson is fitted separately from the leptonic one, and is driven
away from the SM prediction by the excess observed in the 2-jet V$\PH$-tagged category.

\begin{figure*}[!h]
\centering
\includegraphics[width=\cmsFigWidthAlt]{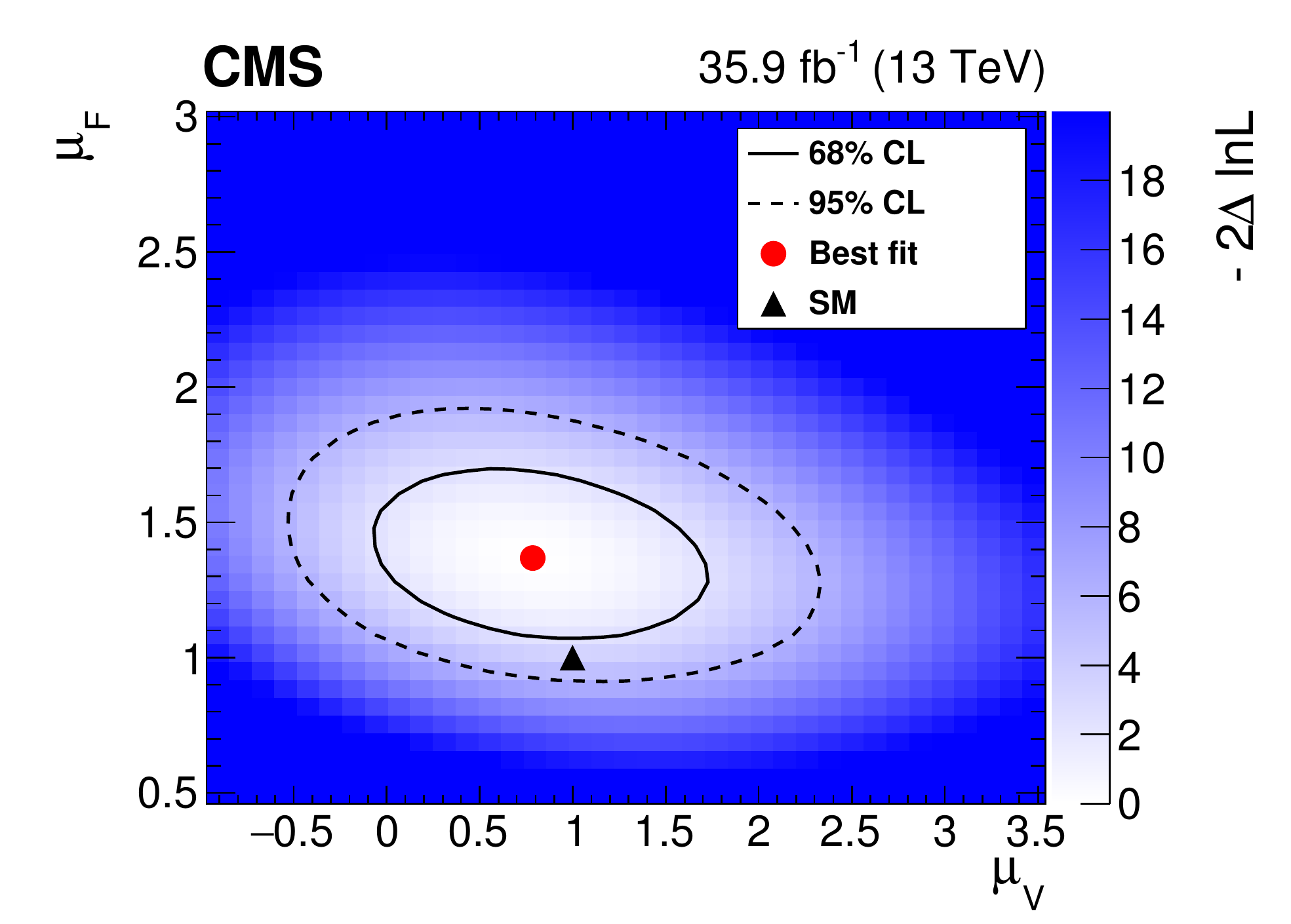}
\includegraphics[width=\cmsFigWidthAlt]{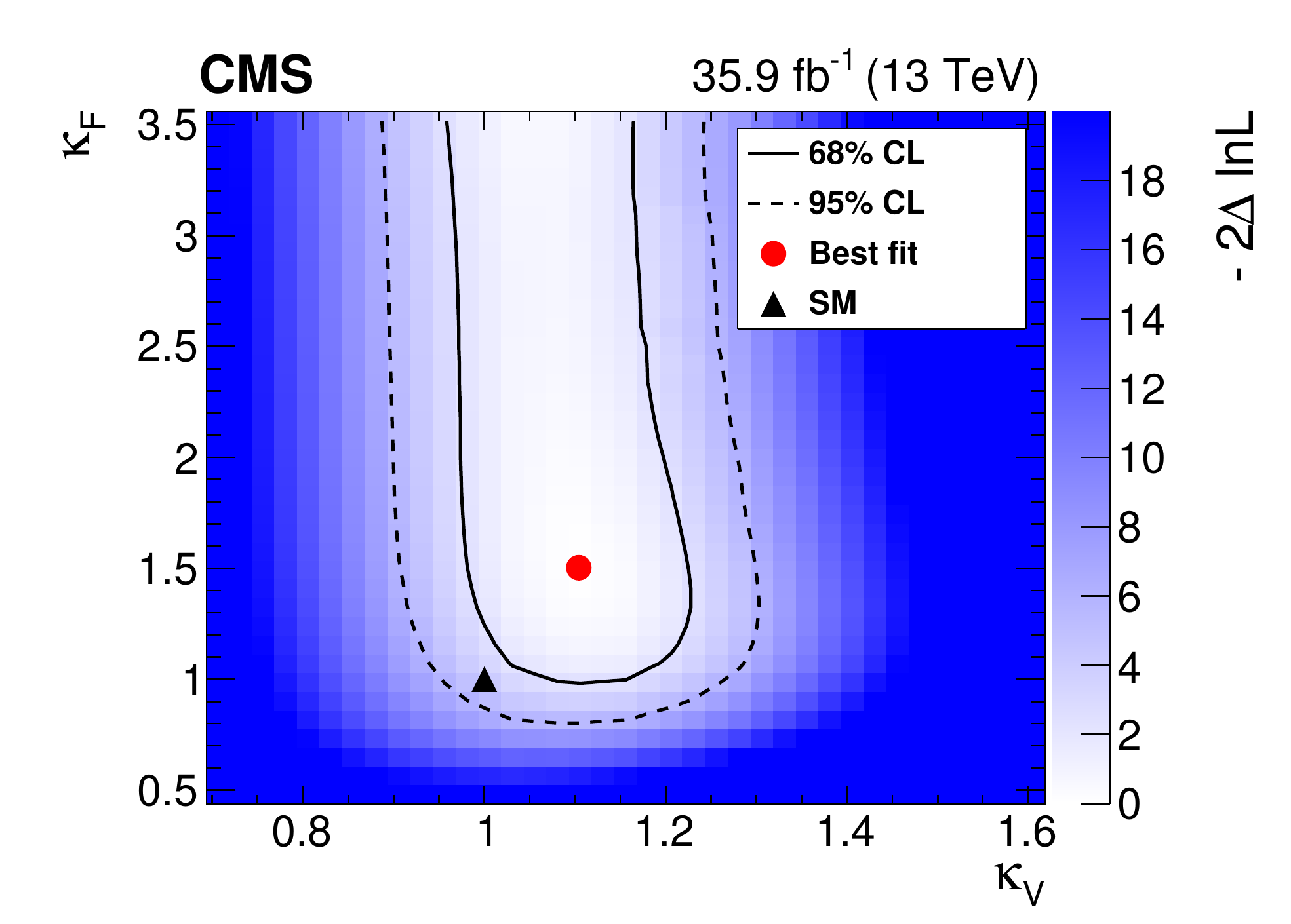}
\caption{
  Two-dimensional likelihood profile as a function of (left) the signal strength
  modifiers associated with either fermion ($\mu{_\mathrm{F}}$) or vector
  boson ($\mu{_\mathrm{V}}$) couplings, and (right) the coupling modifiers associated with either fermion ($\kappa{_\mathrm{F}}$) or vector boson ($\kappa{_\mathrm{V}}$) vertices, using the $\kappa$-framework parametrization.
  The 68\% and 95\% \CL contours are shown as continuous and dashed lines, respectively.
  The red circle represents the best fit value, while the black triangle corresponds to the SM prediction.
}
\label{fig:couplings}
\end{figure*}

\subsection{Higgs boson couplings}

Given its large cross section times branching fraction, the $\PH\to\PW\PW$
channel has the  potential for constraining the Higgs boson couplings to
vector bosons and fermions.
A fit is performed to probe these couplings. One signal strength modifier ($\mu{_\mathrm{F}}$) is used to scale fermion-induced production mechanisms, i.e., $\Pg\Pg\PH$, $\ttbar\PH$, and $\bbbar\PH$, and another one ($\mu{_\mathrm{V}}$) scales the production mechanisms associated with vector bosons, i.e., VBF and V$\PH$.
The two-dimensional likelihood profile is shown in Fig.~\ref{fig:couplings} (left), where the 68\% and 95\% \CL contours in the $(\mu{_\mathrm{F}},\mu{_\mathrm{V}})$ plane are displayed. The best fit values for the signal strength modifiers are $\mu{_\mathrm{F}} = 1.37^{+0.21}_{-0.20}$ and $\mu_\mathrm{V} = 0.78^{+0.60}_{-0.57}$.

The determination of the Higgs boson coupling constants is a way to verify the theoretical predictions and to search for
deviations with respect to the SM expectations. These couplings can be parametrized using two coupling
modifiers associated either with fermion or vector boson vertices, using the
so-called $\kappa$-framework~\cite{deFlorian:2016spz}.
The two coupling modifiers are used to scale the expected product of cross section and branching fraction
to match the observed signal yields in the data, according to the following formula:
\begin{equation}
\sigma \, {{\mathcal B}(\mathrm{X}\to\PH\to\PW\PW)} = \kappa{_\mathrm{i}}^2 \frac{\kappa{_\mathrm{V}}^2}{\kappa{_{\PH}}^2}\sigma{_\mathrm{SM}}\,{{\mathcal B}_\mathrm{SM}(\mathrm{X}\to\PH\to\PW\PW)},
\end{equation}
\noindent where $\kappa{_{\PH}} = \kappa{_{\PH}}(\kappa{_\mathrm{F}},\kappa{_\mathrm{V}})$ is the Higgs boson total width modifier, defined as a function of the two fit parameters $\kappa{_\mathrm{F}}$ and $\kappa{_\mathrm{V}}$. The $\kappa{_\mathrm{i}}$ coupling modifier is equal to $\kappa{_\mathrm{F}}$ for the $\Pg\Pg\PH$, $\ttbar\PH$, and $\bbbar\PH$ production modes, and to $\kappa{_\mathrm{V}}$ for the VBF and V$\PH$ production modes. No processes other than SM ones are considered to contribute to the total width modifier.
The two-dimensional likelihood profile obtained using this approach, and the corresponding 68\% and 95\% \CL contours, are shown in Fig.~\ref{fig:couplings} (right).
The best fit values for the coupling modifiers, obtained with one-dimensional
fits in which the other coupling is profiled, are $\kappa{_\mathrm{F}} = 1.52^{+0.48}_{-0.41}$ and $\kappa{_\mathrm{V}} = 1.10^{+0.08}_{-0.08}$.
The fact that $\kappa{_\mathrm{V}}$ is larger than 1 while the signal strength
modifier $\mu{_\mathrm{V}}$ is below 1 is due to the former being constrained
not only by the production, but also by the decay of the Higgs boson, and thus
being affected by the fact that the global observed signal strength is larger than 1.

\section{Summary\label{conclusion}}

Measurements of the properties of the SM Higgs boson decaying to a $\PW$ boson pair at the LHC have been reported.
The data samples used in the analysis correspond to an integrated luminosity of $35.9\fbinv$ collected by the CMS
detector in proton-proton collisions at $\sqrt{s} = 13\TeV$.

The $\PWp\PWm$ candidates are selected in events with large missing transverse momentum and exactly two, three, or four leptons.
In the case of events with two leptons, different categories are defined according to the lepton pair flavor, $\Pe\Pgm$, $\Pe\Pe$, or $\Pgm\Pgm$.
The analysis has specific categories for gluon fusion production, vector boson fusion, and vector boson associated production, with up to two jets in the final state.

The probability of observing a signal at least as large as the one seen by combining all channels, under the background-only hypothesis,
corresponds to an observed significance of 9.1 standard deviations for $m_{\PH}=125.09\GeV$, to be compared with the expected value of 7.1 standard deviations.
The observed global signal strength modifier is $\sigma/\sigma{_\mathrm{SM}} = \mu = 1.28 ^{+0.18}_{-0.17} = 1.28 \pm 0.10\stat\pm 0.11\syst^{+0.10}_{-0.07}\thy$.
Measurements of the signal strength modifiers associated with the main Higgs boson production mechanisms are also performed, as well as measurements of the Higgs boson couplings to fermions and vector bosons.
The measured Higgs boson production and decay properties are found to be consistent, within their uncertainties, with the SM expectation.

\begin{acknowledgments}
We congratulate our colleagues in the CERN accelerator departments for the
excellent performance of the LHC and thank the technical and administrative
staffs at CERN and at other CMS institutes for their contributions to the
success of the CMS effort. In addition, we gratefully acknowledge the
computing centers and personnel of the Worldwide LHC Computing Grid for delivering so effectively the computing infrastructure essential to our analyses. Finally, we acknowledge the enduring support for the construction and operation of the LHC and the CMS detector provided by the following funding agencies: BMWFW and FWF (Austria); FNRS and FWO (Belgium); CNPq, CAPES, FAPERJ, FAPERGS, and FAPESP (Brazil); MES (Bulgaria); CERN; CAS, MoST, and NSFC (China); COLCIENCIAS (Colombia); MSES and CSF (Croatia); RPF (Cyprus); SENESCYT (Ecuador); MoER, ERC IUT, and ERDF (Estonia); Academy of Finland, MEC, and HIP (Finland); CEA and CNRS/IN2P3 (France); BMBF, DFG, and HGF (Germany); GSRT (Greece); NKFIA (Hungary); DAE and DST (India); IPM (Iran); SFI (Ireland); INFN (Italy); MSIP and NRF (Republic of Korea); LAS (Lithuania); MOE and UM (Malaysia); BUAP, CINVESTAV, CONACYT, LNS, SEP, and UASLP-FAI (Mexico); MBIE (New Zealand); PAEC (Pakistan); MSHE and NSC (Poland); FCT (Portugal); JINR (Dubna); MON, RosAtom, RAS, RFBR, and NRC KI (Russia); MESTD (Serbia); SEIDI, CPAN, PCTI, and FEDER (Spain); Swiss Funding Agencies (Switzerland); MST (Taipei); ThEPCenter, IPST, STAR, and NSTDA (Thailand); TUBITAK and TAEK (Turkey); NASU and SFFR (Ukraine); STFC (United Kingdom); DOE and NSF (USA).

\hyphenation{Rachada-pisek} Individuals have received support from the Marie-Curie program and the European Research Council and Horizon 2020 Grant, contract No. 675440 (European Union); the Leventis Foundation; the A. P. Sloan Foundation; the Alexander von Humboldt Foundation; the Belgian Federal Science Policy Office; the Fonds pour la Formation \`a la Recherche dans l'Industrie et dans l'Agriculture (FRIA-Belgium); the Agentschap voor Innovatie door Wetenschap en Technologie (IWT-Belgium); the F.R.S.-FNRS and FWO (Belgium) under the ``Excellence of Science - EOS" - be.h project n. 30820817; the Ministry of Education, Youth and Sports (MEYS) of the Czech Republic; the Lend\"ulet (``Momentum") Programme and the J\'anos Bolyai Research Scholarship of the Hungarian Academy of Sciences, the New National Excellence Program \'UNKP, the NKFIA research grants 123842, 123959, 124845, 124850 and 125105 (Hungary); the Council of Science and Industrial Research, India; the HOMING PLUS program of the Foundation for Polish Science, cofinanced from European Union, Regional Development Fund, the Mobility Plus program of the Ministry of Science and Higher Education, the National Science Center (Poland), contracts Harmonia 2014/14/M/ST2/00428, Opus 2014/13/B/ST2/02543, 2014/15/B/ST2/03998, and 2015/19/B/ST2/02861, Sonata-bis 2012/07/E/ST2/01406; the National Priorities Research Program by Qatar National Research Fund; the Programa Estatal de Fomento de la Investigaci{\'o}n Cient{\'i}fica y T{\'e}cnica de Excelencia Mar\'{\i}a de Maeztu, grant MDM-2015-0509 and the Programa Severo Ochoa del Principado de Asturias; the Thalis and Aristeia programs cofinanced by EU-ESF and the Greek NSRF; the Rachadapisek Sompot Fund for Postdoctoral Fellowship, Chulalongkorn University and the Chulalongkorn Academic into Its 2nd Century Project Advancement Project (Thailand); the Welch Foundation, contract C-1845; and the Weston Havens Foundation (USA). \end{acknowledgments}

\bibliography{auto_generated}
\cleardoublepage \appendix\section{The CMS Collaboration \label{app:collab}}\begin{sloppypar}\hyphenpenalty=5000\widowpenalty=500\clubpenalty=5000\vskip\cmsinstskip
\textbf{Yerevan Physics Institute, Yerevan, Armenia}\\*[0pt]
A.M.~Sirunyan, A.~Tumasyan
\vskip\cmsinstskip
\textbf{Institut f\"{u}r Hochenergiephysik, Wien, Austria}\\*[0pt]
W.~Adam, F.~Ambrogi, E.~Asilar, T.~Bergauer, J.~Brandstetter, M.~Dragicevic, J.~Er\"{o}, A.~Escalante~Del~Valle, M.~Flechl, R.~Fr\"{u}hwirth\cmsAuthorMark{1}, V.M.~Ghete, J.~Hrubec, M.~Jeitler\cmsAuthorMark{1}, N.~Krammer, I.~Kr\"{a}tschmer, D.~Liko, T.~Madlener, I.~Mikulec, N.~Rad, H.~Rohringer, J.~Schieck\cmsAuthorMark{1}, R.~Sch\"{o}fbeck, M.~Spanring, D.~Spitzbart, A.~Taurok, W.~Waltenberger, J.~Wittmann, C.-E.~Wulz\cmsAuthorMark{1}, M.~Zarucki
% ...
\vskip\cmsinstskip
% ...
1:  Also at Vienna University of Technology, Vienna, Austria\\
% ...
\end{sloppypar}
\end{document}